\documentclass[english]{article}
\usepackage{graphicx}
\usepackage{constants}
\usepackage{color}
\usepackage{amsmath}
\usepackage{amssymb}
\usepackage{amsthm}

\makeatletter

\newcommand{\revise}[1]{#1}

\theoremstyle{plain}
\newtheorem{thm}{\protect\theoremname}
  \theoremstyle{plain}
  \newtheorem{prop}[thm]{\protect\propositionname}
  \theoremstyle{plain}
  \newtheorem{cor}[thm]{\protect\corollaryname}
  \theoremstyle{remark}
  
  \theoremstyle{plain}
  \newtheorem{lem}[thm]{\protect\lemmaname}

\def \l {\left}
\def \r {\right}

\def\alg{\mbox{MC}^{\operatorname{2}}}

\usepackage{babel}
  \providecommand{\corollaryname}{Corollary}
  \providecommand{\lemmaname}{Lemma}
  \providecommand{\propositionname}{Proposition}
  \providecommand{\remarkname}{Remark}
\providecommand{\theoremname}{Theorem}

\begin{document}

\title{$\alg$: A Two-Phase Algorithm for Leveraged Matrix Completion}
\author{Armin Eftekhari, Michael B.\ Wakin, and Rachel A.\ Ward\footnote{AE is with the Alan Turing Institute, London, UK. MBW is with the Department of Electrical Engineering at the Colorado School of Mines, Golden, CO, USA. RW is with the Department of Mathematics at the University of Texas at Austin and a member of the Institute for Computational Engineering and Sciences (ICES), Austin, TX, USA. Email: armin.eftekhari@gmail.com, mwakin@mines.edu, rward@math.utexas.edu. MBW was partially supported by NSF CAREER Grant CCF-$1149225$ and NSF Grant CCF-$1409258$. RW was partially funded by NSF CAREER Grant  CCF-$1255631$.}}
\maketitle

\begin{abstract}
{Leverage scores, loosely speaking, reflect the importance of the rows and columns of a matrix. Ideally, given the leverage scores of a rank-$r$ matrix $M\in\mathbb{R}^{n\times n}$, that matrix can be reliably completed from just $O(rn\log^{2}n)$ samples if the samples are chosen randomly from a nonuniform distribution induced by the leverage scores. In practice, however, the leverage scores are often unknown \emph{a priori}. As such, the sample complexity in uniform matrix completion---using uniform random sampling---increases to $O(\eta(M)\cdot rn\log^{2}n)$, where $\eta(M)$ is the largest leverage score of $M$. In this paper, we propose a two-phase algorithm called $\alg$ for matrix completion: in the first phase, the leverage scores are estimated based on uniform random samples, and then in the second phase the matrix is resampled nonuniformly based on the estimated leverage scores and then completed. \revise{For well-conditioned matrices, the total sample complexity of $\alg$ is no worse than uniform matrix completion, and for certain classes of well-conditioned matrices---namely, reasonably coherent matrices whose leverage scores exhibit mild decay---$\alg$ requires substantially fewer samples. Numerical simulations suggest that the algorithm outperforms uniform matrix completion in a broad class of matrices, and in particular, is much less sensitive to the condition number than our theory currently requires.}}%{Matrix completion, leverage scores, coherence, nuclear norm minimization}
%\\
%2000 Math Subject Classification: 34K30, 35K57, 35Q80,  92D25
\end{abstract}

\section{Introduction}

\emph{Matrix completion} is commonly defined as the problem of recovering a  low-rank matrix $M\in\mathbb{R}^{n_{1}\times n_{2}}$
from a fraction of its entries, observed on an often
random index set \cite{candes2009exact,eriksson2012high,recht2011simpler}.  To be concrete,  let  $n_{1}=n_{2}=n$ and set
$r=\mbox{rank}(M)$ for short. Also let $M=U \Sigma V^*$ be the ``skinny" singular value decomposition (SVD) of $M$, where $U,V \in\mathbb{R}^{n\times r}$ have orthonormal columns and the diagonal matrix $\Sigma\in\mathbb{R}^{r\times r}$ contains the singular values of $M$.

In \emph{uniform} low-rank matrix completion (UMC), each entry of $M$ is observed with a probability of $p\in(0,1]$ so that, in expectation, $pn^2$ entries of $M$ are revealed.  As reviewed in more detail in Section \ref{sec: A brief review}, from these uniform samples, $M$ can be successfully reconstructed
 (via convex programming, for example) provided that
 \begin{equation}
  \eta(M)\cdot\frac{r\log^2 n}{n} \lesssim p \le 1, \label{eq:uniform intro}
 \end{equation}
where $\eta(M)$, the \emph{coherence} of $M$,  in a sense measures how ``diffuse''  $M$ is. Note that, throughout, we often use $\lesssim$ to simplify the presentation by omitting universal constant factors. Above, the dependence of $p$ on $r$ and $n$ is optimal in worst-case complexity, up to a logarithmic factor, and also
\begin{equation}
\eta(M) := \frac{n}{r} \l( \l\| U \r\|_{2\rightarrow\infty}^2 \vee  \l\| V\r\|_{2\rightarrow\infty}^2 \r)
=
\frac{n}{r}  \l( \max_{i\in[1:n]} \l\|U[i,:]\r\|_2^2  \vee  \max_{j\in[1:n]} \l\|V[j,:]\r\|_2^2 \r),
\end{equation}
with $U[i,:]$ and $V[j,:]$ standing for the  corresponding rows of $U$ and $V$, respectively. We also use the conventions $a\vee b=\max[a,b]$ and $[c:d]=\{c,c+1,\cdots,d\}$ for integers $c\le d$. One may verify that $\eta(M)\in[1,\frac{n}{r}]$. It is also common to say that $M$ is \emph{coherent} (\emph{incoherent}) when $\eta(M)$ is very large (small). Loosely  speaking, a coherent matrix is ``spiky'' whereas an incoherent matrix is ``diffuse'' with respect to the distribution of the magnitudes of its entries. For example, if $M[1,1]=1$ is the only nonzero entry of $M$, then $M$ is extremely coherent since $\eta(M)=\frac{n}{r}$.

Roughly speaking, in UMC,  we can expect to successfully recover $M$ from  $O(\eta(M)\cdot rn\log^{2}n) $
uniform samples. In particular, when $M$ is {incoherent}, say $\eta(M)\approx 1$,  then $M$ can be completed from  $O(rn\log^2n )\ll n^2$ uniform samples. In contrast, when $M$ is coherent, say $\eta(M)\approx \frac{n}{r}$,  then one needs to observe nearly all entries of $M$.  For instance, if $M[1,1]=1$ is the only
nonzero entry of $M$, then uniform sampling will collect $M[1,1]$ with a probability of $p$. So, unless $p\approx 1$,  $M[1,1]$ is not observed and successful reconstruction of this coherent matrix from uniform samples is highly unlikely.

The poor performance of UMC in completing coherent matrices can be remedied by means of  \emph{leveraged} (rather than uniform) sampling \cite{chen2013completing}. In the example above, $M[1,1]$ is by far the most
important entry of $M$. Therefore, a better sampling strategy might be to measure $M[1,1]$ with more likelihood than the rest of the entries.
More generally, the importance of the rows and columns of a rank-$r$ matrix $M\in\mathbb{R}^{n\times n}$ are often measured by its \emph{leverage scores} defined as
\begin{equation}
\mu_{i}(M):=\frac{n}{r}\left\Vert U[i,:]\right\Vert _{2}^{2},\qquad\nu_{j}(M):=\frac{n}{r}\left\Vert V[j,:]\right\Vert _{2}^{2},
\qquad  i,j\in [1:n].
\label{eq:leverages def in setup}
\end{equation}
It is easily verified that $\mu_i(M),\nu_j(M)\in [0,\frac{n}{r}]$ for all $i,j\in[1:n]$ and that the coherence $\eta(M)$ is simply the largest leverage score of $M$. Moreover,
\begin{equation}
\sum_{i=1}^{n}\mu_{i}(M)=\sum_{j=1}^{n}\nu_{j}(M)=n,
\label{eq:sum of mus nus}
\end{equation}
since $U^{*}U=V^{*}V=I_{r}$, with  $I_{r}\in\mathbb{R}^{r\times r}$ being the identity matrix. Naturally, we might consider $\mu_i(M)+\nu_j(M)$ as an indicator of the importance of $M[i,j]$. In the example where  $M[1,1]=1$ is the only nonzero entry of $M$, $\mu_1(M)+\nu_1(M)= 2n$, whereas $\mu_i(M)+\nu_j(M)= 0$ for every $i,j>1$,  suggesting the importance of the first row and column of $M$.

\emph{If }the leverage scores of $M$ were known in advance, a good sampling strategy would be to  measure the entries of $M$ according to their
importance $\mu_i(M)+\nu_j(M)$. More specifically, in \emph{leveraged} low-rank matrix completion (LMC), $M$ can be recovered (via convex programming, for instance)  provided that each entry $M[i,j]$ is observed with a probability of $P[i,j]$ that satisfies
\begin{equation}
\l(\mu_i(M)+\nu_j(M)\r) \frac{r\log^2n}{n} \lesssim P[i,j] \le 1, \qquad \forall i,j\in[1:n].
\end{equation}
That is,  we can expect to recover $M$ from
\begin{equation}
 \sum_{i,j=1}^n P[i,j] = \sum_{i,j} \l( \mu_i(M)+\nu_j(M) \r)
\cdot \frac{r\log^2n}{n}
= O\l( rn\log^2 n \r)
\qquad \mbox{(see \eqref{eq:sum of mus nus})}
\end{equation}
entries, as opposed to $O(\eta(M)\cdot rn\log^{2}n)$ uniform samples required in
UMC,  thereby removing any  dependence on coherence, and improving the sample complexity of low-rank matrix completion by up to a factor of $\frac{n}{r}$.

\subsection{Our contributions}
In practice, the leverage scores of $M$ are often unknown a priori, suggesting the need for a matrix completion scheme that would improve over UMC, particularly in completing coherent matrices, and yet not require much prior knowledge about $M$. In this paper, we propose
a {two-phase algorithm }for matrix completion---dubbed $\alg$---which
\begin{enumerate}
\item first estimates the relatively large leverage scores of $M$ from uniform or ``oblivious" samples,
\item draws a second batch of samples from a weighted distribution according to the estimated leverage scores from the first phase, and finally
\item completes $M$ using both batches of samples, using for example convex optimization.
\end{enumerate}
The prototype algorithm $\alg$ is developed in Section \ref{sec:A-Two-Phase-Algorithm} and summarized in Figure \ref{fig:MC:-A-two-phase}.

\revise{Unlike LMC, $\alg$ requires little prior knowledge about $M$ and yet substantially improves over  UMC when, loosely speaking, $M$ is well-conditioned and \emph{reasonably coherent} ($1 \ll \eta(M) \ll \frac{n}{r}$ ).} Our main sample complexity result is stated in Theorem \ref{thm:main} below.  The main point is that the sample complexity can be improved for such matrices because one only needs rough estimates of the \emph{largest} leverage scores of a low-rank matrix to apply the LMC theory, and one does not need sophisticated concentration inequalities to get such estimates---Chebyshev's inequality suffices.  Thus, in general, fewer uniformly distributed entrywise samples are needed to obtain rough estimates for the largest leverage scores of a low-rank matrix than are needed to complete such a matrix entirely.  Lemma \ref{lemma:estLev}, as a byproduct of our analysis, provides a new bound on the sample complexity for estimating a subset of the leverage scores of a low-rank matrix. This is interesting in its own right, given the additive nature of most of the available bounds in the literature of numerical linear algebra.

\revise{For well-conditioned matrices, the total sample complexity of $\alg$ is no worse than uniform matrix completion, and for certain classes of well-conditioned matrices---namely, reasonably coherent matrices whose leverage scores exhibit mild decay---$\alg$ requires substantially fewer samples. It is worth mentioning that the polynomial dependence on the condition number in the number of samples in our main result appears to be an artifact of the proof techniques; in numerical simulations, $\alg$ improves over UMC for a broad class of matrices and its performance degrades only mildly as the condition number $\kappa$ increases.} A variant of $\alg$ first appeared in \cite{chen2013completing} where it was shown to outperform UMC in numerical simulations.  Our main contribution in this paper is in carefully studying the performance of two-phase sampling as suggested there.

For clarity of exposition, we only consider the noiseless, exact low-rank case in this work.  It is not difficult to extend the results here to matrices which are nearly low-rank, and to observations which have a small amount of noise, but this requires a more technical analysis which detracts from the main message of this paper.  In future work, we hope to derive a simpler and stronger analysis for this more general setting.

\paragraph{Organization}
The rest of this paper is organized as follows. After a brief review, in Section \ref{sec: A brief review}, of the relevant concepts in matrix completion,  $\alg$ is developed in Section \ref{sec:A-Two-Phase-Algorithm} and detailed in Figure \ref{fig:MC:-A-two-phase}. The accompanying  theoretical guarantees are given in \revise{Theorem~\ref{thm:main}, Corollary~\ref{cor:main}, and Corollary~\ref{cor:second}.} Without being exhaustive, Section \ref{sec:Numerical-Simulations}  compares $\alg$ and UMC numerically. Related work is discussed in Section  \ref{sec:Related-Work}.

\section{Matrix Completion: A Brief Review}
\label{sec: A brief review}

Consider a rank-$r$ matrix $M\in\mathbb{R}^{n\times n}$ with ``skinny" SVD decomposition  $M=U\Sigma V^{*}$. Here,
$U,V\in\mathbb{R}^{n\times r}$ consist of orthonormal columns, and the diagonal
matrix $\Sigma\in\mathbb{R}^{r\times r}$ collects the non-zero singular values
of $M$,  namely $\sigma_{1}(M)\ge\sigma_{2}(M)\ge\cdots\ge\sigma_{r}(M)$, in non-increasing order (ties are broken arbitrarily).

Central to this work, the row and column \emph{leverage scores} of $M$ are defined \revise{as in~\eqref{eq:leverages def in setup}.} It is easy to see that $\mu_i(M),\nu_j(M)\in [0,\frac{n}{r}]$ for every $i,j\in[1:n]$ \revise{and that each set of leverage scores sums to $n$ (see~\eqref{eq:sum of mus nus}),} so that $\{\mu_i(M)/n \}_i,\{\nu_j(M)/n \}_j$ might be interpreted as probability distributions on the rows and columns of $M$, respectively.
In a sense, leverage scores capture the importance of the rows and columns of $M$.
The {coherence} of $M$ is set to be the largest leverage score, namely
\begin{equation}
\eta(M)=\max_{i\in [1:n]}\mu_{i}(M)\vee\max_{j\in [1:n]}\nu_{j}(M) \in[1,\frac{n}{r}].\label{eq:def of coh}
\end{equation}
We may only access $M$ through entrywise measurements $M[i,j]$ and the goal is to recover $M$ from as few such measurements as possible.
For purposes of analysis, we assume that the entries of $M$ are revealed randomly.  In the uniform matrix completion (UMC) set-up, entries are revealed independently and with equal probability $p \in (0,1]$. The following result is a corollary of Theorem 2 in \cite{chen2013completing} (see also \cite{chen2015incoherence}), improving over results in previous works \cite{candes2009exact,eriksson2012high,recht2011simpler}:
\begin{prop}
\label{prop:vanilla MC}
\textbf{\emph{[Uniform matrix completion (UMC)]}}
Fix a rank-$r$ matrix $M\in\mathbb{R}^{n\times n}$ and probability $p\in(0,1]$.  For each $(i,j) \in [1:n] \times [1:n]$, the entry $M[i,j]$ is independently observed with probability $p$.  Let $\Omega$ denote the set of observed indices.  Let $\widehat{M}\in\mathbb{R}^{n\times n}$ be a solution of
\begin{equation}
\begin{cases}
\min_{X}\|X\|_{*},\\
X[i,j] = M[i,j] , \quad (i,j) \in \Omega
\end{cases}
\label{eq:standard MC thm}
\end{equation}
where $\|\cdot\|_*$ denotes the nuclear norm of a matrix (sum of its singular values). Then, except with a probability of $\frac{1}{n^{10}}$, we have exact low-rank matrix recovery, $\widehat{M} = M$,
provided that
\begin{equation}
\revise{C_1}\eta(M)\cdot\frac{r\log^{2}n}{n} \leq p\le1. \label{eq:to be checked 2}
\end{equation}
where \revise{$C_1 > 0$} is a universal constant, independent of all dimensions and parameters.
\end{prop}
The expected size of $\Omega$ under this set-up is $p n^2 \gtrsim \eta(M) \cdot n r\log^{2}n$, and a straightforward consequence of Hoeffding's inequality gives that UMC completes a rank-$r$ matrix $M\in\mathbb{R}^ {n\times n}$ with overwhelmingly high probability from $O(\eta(M)\cdot  rn\log^2n)$ of its entries, uniformly at random.  UMC is particularly powerful when $M$ is \emph{incoherent}, $\eta(M)\approx 1$, in which case UMC requires only $O(rn\log^2 n)$ uniform samples. For more \emph{coherent} matrices, UMC requires increasingly more uniform samples. At worst, when $\eta(M)=\frac{n}{r}$, we must observe nearly all entries of $M$.

We remark that there are alternatives to Program \eqref{eq:standard MC thm} for matrix completion; see for example \cite{keshavan2010matrix,jain2013low,fornasier2011low,cai2010singular}, among many other algorithms.

The poor performance of UMC in completing coherent matrices  is tied to the uniform sampling strategy. If the leverage scores of $M$ were known in advance,
a better sampling strategy would be to measure important entries of $M$ (namely those corresponding to large leverage scores) with more likelihood, rather than sampling $M$ uniformly at random. Indeed, \emph{leveraged} sampling generalizes the results of UMC to the setting where the leverage scores are not uniform, and leads to substantial improvement over UMC, as we next describe. In {leveraged matrix completion} (LMC), given the knowledge of the leverage scores $\mu_i(M)$ and $\nu_j(M)$, $i,j \in [1:n]$ (or upper bounds on these quantities), we recover $M$ from entries drawn from a weighted distribution biased towards rows and columns with large leverage scores.  The following result is a reformulation of Theorem 2 in \cite{chen2013completing}.
\begin{prop}
\label{thm:gen Rachel's thm}
\emph{\textbf{[Leveraged matrix completion]}}
Fix a rank-$r$ matrix $M\in\mathbb{R}^{n\times n}$
and matrix of probabilities $P\in[0,1]^{n\times n}$.  For each $(i,j) \in [1:n] \times [1:n]$, the entry $M[i,j]$ is independently observed with probability $P[i,j]$.  Let $\Omega$ denote the set of observed indices.  Let $\widehat{M}\in\mathbb{R}^{n\times n}$ be a solution of
\begin{equation}
\begin{cases}
\min_{X}\|X\|_{*},\\
X[i,j] = M[i,j] , \quad (i,j) \in \Omega.
\end{cases}
\end{equation}
Then, except with a probability of $\frac{1}{n^{10}}$, we have exact low-rank matrix recovery, $\widehat{M} = M,$ provided that
\begin{equation}
\label{eq:recguarantee}
\revise{C_2} \left(\mu_{i}(M)+\nu_{j}(M)\right)\frac{r\log^{2}n}{n} \leq P[i,j]\le1,\qquad i,j\in[1:n],
\end{equation}
\revise{and that
\begin{equation}
1/n^4 \leq C_2 \left(\mu_{i}(M)+\nu_{j}(M)\right)\frac{r\log^{2}n}{n} ,\qquad i,j\in[1:n],
\label{eq:n4lowerbound}
\end{equation}}
where \revise{$C_2$} is a universal constant, independent of all dimensions and parameters.
\end{prop}
\noindent Noting the normalization
\begin{equation}
\sum_{i,j=1}^n (\mu_i(M) + \nu_j(M) ) = 2n^2,
\end{equation}
it follows that LMC completes a rank-$r$ matrix $M\in\mathbb{R}^{n\times n}$ from $O(rn\log^2 n)$ entrywise observations, \emph{indepedent} of its coherence.  Thus, while in theory this can result in a factor of $\frac{n}{r}$ improvement in sample complexity over UMC, in practice, the leverage scores of $M$ are often not known a priori; this impedes the practical implementation of LMC. We set out to address this problem next.

\section{$\alg$: A Two-phase Algorithm for Low-rank Matrix Completion}
\label{sec:A-Two-Phase-Algorithm}

So far, we  reviewed uniform and leveraged matrix completion, and explained how the lack of \emph{a priori} knowledge about the leverage scores impedes the implementation of leveraged sampling in practice. To resolve this issue, consider a two-phase algorithm which, in Phase 1, estimates the leverage scores from a small number of uniform samples, and in Phase 2, uses these estimated leverage scores for leveraged matrix completion. The idea for such a two-phase algorithm was presented in \cite{chen2013completing}, albeit without any theoretical guarantees of being superior to uniform matrix completion. Clearly, for extremely coherent cases such as matrices having only one non-zero entry, nothing can be done without more prior information.  The main insight of this paper is that, for many moderately coherent low-rank matrices, two-phase adaptive sampling can provably complete low-rank matrices using fewer total samples than if all the samples were drawn uniformly. A key insight in our analysis is that, in Proposition \ref{thm:gen Rachel's thm} on leveraged matrix completion, in equation \eqref{eq:recguarantee}, one only needs \emph{bounds} on \emph{sufficiently large} leverage scores of the underlying matrix; small leverage scores are bounded automatically using the uniform samples obtained in Phase 1 and no further samples on this portion of the matrix are needed.  The two-phase algorithm is described in Figure \ref{fig:MC:-A-two-phase}.

\begin{figure}[p]
\begin{center}%
\fbox{\begin{minipage}[t]{1\columnwidth}%
\textbf{Input: }
\begin{itemize}
\item Rank $r$ and condition number $\kappa = \sigma_1 / \sigma_r$ of rank-$r$ matrix $M \in\mathbb{R}^{n\times n}$, and Phase 1 measurement budget of $p\in(0,1]$.
\item Access to entrywise sampling of $M \in\mathbb{R}^{n\times n}$.
\end{itemize}
\textbf{Output:}
\begin{itemize}
\item Estimate $\widehat{M}\in\mathbb{R}^{n\times n}$ of $M$.
\end{itemize}
\textbf{Body:}
\begin{enumerate}

\item \textbf{(Phase 1: Uniform sampling)} Observe each entry of $M$ independently with a probability of $p$: let  $Y\in\mathbb{R}^{n\times n}$  store the measurements, filled with zeros elsewhere. Let $\Omega\subseteq [1:n]^2$ be the corresponding index set over which $M$ is observed.
\item \textbf{(Estimate the leverage scores)} Set
\[
\widehat{\mu}_{i}\leftarrow  \frac{ n \kappa^2  \|  Y[i,:] \|_F^2}{ \| Y \|_F^2 } ,\qquad i\in[1:n],
\]
\begin{equation}
\widehat{\nu}_{j}\leftarrow  \frac{n \kappa^2 \|  Y[:,j] \|_F^2}{ \| Y \|_F^2 }, \qquad j\in[1:n].
\label{eq:mu bar alg}
\end{equation}

\item \textbf{(Phase 2: Leveraged sampling)} Set
\begin{equation}
P[i,j]\leftarrow  \min\{1, \frac{3 \revise{C_2} r \log^2(n)}{n} \left(\widehat{\mu}_{i}+\widehat{\nu}_{j}\right)\},  \qquad i,j\in[1:n],
\label{eq:def of Pij in alg}
\end{equation}
where \revise{$C_2$} is the universal constant from Proposition \ref{thm:gen Rachel's thm}.  Then, observe the $[i,j]$th entry of $M$  with a probability of $P[i,j] + p$, for each $i,j\in[1:n]$. Add the resulting index set to $\Omega$.
\item \textbf{(Matrix completion)} Let $\widehat{M}$ be a solution
of the program
\begin{equation}
\begin{cases}
\min_{X}\|X\|_{*},\\
X[i,j] = M[i,j] , \quad (i,j) \in \Omega.
\end{cases}
\label{eq:pr Alg I}
\end{equation}
\end{enumerate}%
\end{minipage}}\end{center}

\protect\caption{$\alg$: A two-phase algorithm for leveraged matrix completion.\label{fig:MC:-A-two-phase}}
\end{figure}

\begin{subsection}{Theoretical guarantees}
\label{sec:theoryresults}

\revise{In this section, we present our theoretical guarantees on the performance of $\alg$; Section~\ref{sec:proofs} contains all proofs.}

The intuition for why the two-phase matrix completion algorithm has lower sample complexity than uniform matrix completion in many instances is that estimating a small number of leverage scores of a low-rank requires fewer uniform entrywise samples than completing the low-rank entirely.  In particular, for the former task, we do not need strong concentration and can use very basic inequalities such as Chebyshev's inequality, as opposed to matrix  Bernstein inequalities for the latter task.  Chebyshev's inequality does not depend on an $L_{\infty}$ uniform upper bound on the input random variables, and as such the following lemma, which is a key ingredient in the analysis of our main theorem, has weaker dependence on the matrix coherence compared to uniform matrix completion results.

\begin{lem}[Estimating \revise{the largest} leverage scores from uniform samples]
\label{lemma:estLev}
 Consider the notation of $\alg$ in Figure~\ref{fig:MC:-A-two-phase}.  Let $\mu_{(1)} \geq \mu_{(2)} \geq \dots \geq \mu_{(n)}$ and $\nu_{(1)} \geq \nu_{(2)} \geq \dots \geq \nu_{(n)}$  be the row (resp.\ column) leverage scores of the rank-$r$ matrix $M$, arranged in decreasing order.  Let $\kappa = \kappa(M) = \sigma_1 / \sigma_r$ be the condition number of $M$, and fix integers $L, d_1, d_2 \in [1:n],$ and fix $\tau \in [0,1/3]$.  Provided that the Phase 1 sampling probability $p \in (0,1]$ is sufficiently large that
$$
pn^2 \geq   16 L \tau^{-1}  \kappa^4 r n  \left(  \frac{r}{n} \sum_{j=1}^{d_1} \nu_{(j)}^2 + \nu_{(d_1+1)} +  \frac{r}{n} \sum_{i=1}^{d_2} \mu_{(i)}^2   + \mu_{(d_2+1)} \right) ,
$$
it holds with probability at least $1-\tau$ that
\begin{align}
\label{eq:ls_est}
\frac{1}{3} \mu_{(i)} &\leq \widehat{\mu}_{(i)} \leq 3 \kappa^4 \mu_{(i)}  , \quad \quad  i \in [1:L] \nonumber \\
\frac{1}{3}  \nu_{(j)}  &\leq \widehat{\nu}_{(j)} \leq 3 \kappa^4 \nu_{(j)},  \quad \quad j \in [1:L],
\end{align}
where $\widehat{\mu}_{(i)}$ and $\widehat{\nu}_{(j)}$ denote the corresponding estimated leverage scores defined in~\eqref{eq:mu bar alg}.
\end{lem}

With Lemma \ref{lemma:estLev} in hand, we can now state and prove the main recovery result for the two-phase algorithm.

\begin{thm}[Main result]
\label{thm:main}
\revise{Let $\mu_{(1)} \geq \mu_{(2)} \geq \dots \geq \mu_{(n)}$ and $\nu_{(1)} \geq \nu_{(2)} \geq \dots \geq \nu_{(n)}$  be the row (resp.\ column) leverage scores \eqref{eq:leverages def in setup} of a rank-$r$ matrix $M = (M[i,j])$ in decreasing order. Suppose that~\eqref{eq:n4lowerbound} holds for all $i,j \in [1:n]$. Let $\kappa = \kappa(M) = \sigma_1 / \sigma_r$ be the condition number of $M$. Fix $\tau \in [0,1/3]$. If
\begin{align}
  \label{rec:newnew}
p & \geq C_3 \cdot \frac{r}{n} \cdot \min_{L \in [0:n]} \left( \max\{ L \tau^{-1}  \kappa^4, \log^2(n) \} \left(  \frac{r}{n} \sum_{j=1}^{L} \nu_{(j)}^2 + \nu_{(L+1)} +  \frac{r}{n} \sum_{i=1}^{L} \mu_{(i)}^2   + \mu_{(L+1)} \right)\right),
\end{align}
then with probability exceeding $1 - (\tau + \frac{1}{n^{10}})$ the two-phase algorithm $\alg$ in Figure~\ref{fig:MC:-A-two-phase} recovers $M$ as $\widehat{M} = M$
using an expected total number of samples
\begin{equation}\label{eq:omegasize}
\mathbb{E} | \Omega | \leq 2 p n^2 + 6 \revise{C_2} r n \kappa^2 \log^2(n).
\end{equation}
Above, $C_1$ is the universal constant from Proposition~\ref{prop:vanilla MC}, $C_2$ is the universal constant from Proposition~\ref{thm:gen Rachel's thm}, and $C_3 = \max\{16,C_1,C_2\}$.}
 \end{thm}

\revise{In order to interpret this theorem, we provide a comparison with the corresponding guarantee for UMC. First, note that \eqref{rec:newnew} actually comprises $n + 1$ sufficient conditions (corresponding to each possible value of $L \in [0:n]$ in the minimization term), and for the theorem to apply, the condition \eqref{rec:newnew} need only be satisfied for one value of $L \in [0:n]$. Substituting $L = 0$ into the right-hand side of \eqref{rec:newnew} gives the sufficient condition
\[
p \geq C_3 \cdot \frac{r}{n} \log^2(n) \left(  \nu_{(1)} + \mu_{(1)} \right),
\]
which (up to a constant) matches the classical sufficient condition~\eqref{eq:to be checked 2} on the uniform sampling probability $p$ for exact matrix recovery with UMC listed in Proposition~\ref{prop:vanilla MC}. Thus, for any matrix $M$, the sufficient condition on the uniform sampling probability $p$ for $\alg$ is {\em no worse} (up to a constant) than the classical sufficient condition on $p$ for UMC. For some classes of matrices $M$, the minimizer of the right-hand side of \eqref{rec:newnew} will in fact occur for some $L > 0$. For these matrices, the sufficient condition on the uniform sampling probability $p$ for $\alg$ is {\em better} than the classical sufficient condition on $p$ for UMC. Corollaries~\ref{cor:main} and~\ref{cor:second} below give examples of such classes of matrices.
}

\revise{We also note that a proper comparison with UMC requires not only discussing the uniform sampling probability $p$ but also accounting for the {\em total} number of samples used in each algorithm. Neglecting constants, for a given value of $p$, the total number of samples in UMC scales with $pn^2$, while the total number of samples in $\alg$ scales with $p n^2 + r n \kappa^2 \log^2(n)$ (see~\eqref{eq:omegasize}). If $p$ were chosen according to the classical sufficient condition~\eqref{eq:to be checked 2} for UMC, then $pn^2$ scales with $\eta(M) r n \log^{2}n$. Thus, for well-conditioned matrices (where $\kappa^2 \lesssim \eta(M)$), the total sample complexity of $\alg$ will never exceed (up to a constant) the total sample complexity of UMC.}

\revise{In order to get a sense of when the guarantees for $\alg$ can actually improve upon those for UMC, recall that $\eta(M) = \max\{ \mu_{(1)}, \nu_{(1)} \} \in [1, \frac{n}{r}]$, with $\eta = \frac{n}{r}$ corresponding to ``as coherent as possible'' and $\eta = 1$ being ``as incoherent as possible''.  In the completely coherent case, no algorithm can get around observing most of the matrix entries in order to complete the matrix. Indeed, when $\eta = \frac{n}{r}$, both \eqref{eq:to be checked 2} and \eqref{rec:newnew} require $p = O(1)$. In the completely incoherent case, both \eqref{eq:to be checked 2} and \eqref{rec:newnew} require $p = O(r n^{-1} \log^2{n})$. An opportunity for improvement, however, comes in the intermediate ``somewhat coherent'' case where the maximal leverage score is on the order of $\sqrt{n/r}$ and only a small number of the remaining leverage scores are of the same order of magnitude as the largest. A particular class of such matrices would be those whose leverage scores exhibit power-law decay.  Such matrices were considered in \cite{papailiopoulos2014provable} as a model in providing theoretical guarantees with deterministic leverage score sampling for the column subset selection problem.  In the same paper, empirical evidence was provided that such decay is abundant in real-world settings. For such matrices which are well-conditioned, the sample complexity bound is significantly better than the bound available using standard uniform sampling. This is seen by choosing $L$ in~\eqref{rec:newnew} to be equal to the number of large leverage scores and is quantified in the following result.}

\begin{cor}
\label{cor:main}
\revise{Suppose the rank-$r$ matrix $M$ has coherence $\eta = \max\{ \mu_{(1)}, \nu_{(1)} \} \geq \sqrt{n/r}.$ Moreover, for fixed $T > 0$, suppose its largest leverage scores admit a power-law decay:  for $i,j \leq \lceil{ \eta^{\frac{1}{1+T}} \rceil},$
\begin{align}
\mu_{(i)} &\leq  \mu_{(1)} i^{-(1+T)}, \nonumber \\
\nu_{(j)} &\leq \nu_{(1)} j^{-(1+T)}.
\end{align}
Assuming also the mild condition that~\eqref{eq:n4lowerbound} holds for all $i,j \in [1:n]$, the two-phase algorithm $\alg$ with sampling probability
$$p = 8C_3 \tau^{-1} \kappa^4 \frac{r^2}{n^2} \eta^{\frac{3+2T}{1+T}}
$$
recovers $M$ as $\widehat{M} = M$ with probability exceeding $1 - (\tau + \frac{1}{n^{10}})$ from an expected total number of samples
\begin{align}
\label{power:complexity}
\mathbb{E} | \Omega | &\leq 2 p n^2 + 6C_2 r n \kappa^2 \log^2(n) \nonumber \\
 &\lesssim \tau^{-1} \kappa^4 r^2 \eta^{\frac{3+2T}{1+T}}  + r n \kappa^2 \log^2(n).
\end{align}
Above, $C_2$ is the universal constant from Proposition \ref{thm:gen Rachel's thm}, and $C_3$ is the universal constant from Theorem~\ref{thm:main}.}
\end{cor}

 \noindent Some remarks are in order.

  \begin{itemize}
  \item  The number of samples in Corollary \ref{cor:main} is smaller than the sample complexity for UMC for well-conditioned matrices.  Ideally, when $\kappa = O(1)$, the number of samples in Corollary \ref{cor:main} reduces to
  $$
  | \Omega | = O(\max\{ rn \log^2(n), r^2 \eta^{\frac{3+2T}{1+T}} \}),
  $$
which is significantly smaller than $O(\eta rn \log^2(n))$ for reasonably coherent matrices, namely, those satisfying $\eta \leq ( \frac{n}{r} )^{\frac{1+T}{2+T}}$.   For example, taking $\eta = \sqrt{n/r}$ and $T = 1/2$ gives
a sample complexity for the two-phase algorithm of
   $$
  | \Omega | = O(\max\{ rn \log^2(n), n^{4/3} r^{2/3} \}),
  $$
which is significantly smaller than the bound of $| \Omega | = \eta r n \log^2(n) = n^{3/2} r^{1/2} \log^2(n)$ for UMC.

\item Note that the two-phase algorithm requires setting a parameter which depends on knowing the condition number $\kappa$ a priori, and that the  improved bound for matrices whose leverage scores exhibit power-law decay holds for a particular choice of the sampling probability $p$ which requires moreover knowledge of the power-law decay. It would be interesting in future work to examine whether these assumption can be removed. In Section~\ref{sec:expsetup} we describe a practical implementation of $\alg$ in which knowledge of $\kappa$ is not required.

\item The fourth-order dependence on $\kappa$ in the sample complexity bounds for $\alg$ is likely pessimistic. In experiments described in Sections~\ref{sec:powerdiscussion} and~\ref{sec:blockdiscussion}, we see at worst a quadratic scaling of the leverage score estimation errors as a function of $\kappa$. This raises the question of whether the $\kappa^4$ term in \revise{bounds such as~\eqref{rec:newnew}} could ultimately be improved to $\kappa^2$. This dependence could perhaps be improved even more using a more sophisticated algorithm along the lines of what is done in the paper~\cite{hardt2014fast}, but likely at the expense of a worse dependence on the rank $r$. Finally, the ultimate matrix recovery performance (not leverage score estimation error) in simulations is sometimes robust to condition number (see Section~\ref{sec:powerdiscussion}), though not always (See Section~\ref{sec:blockdiscussion}).

\end{itemize}

  As a second corollary of the main result, we note that under mild conditions on the leverage scores, the $\log(n)$ factors in the sample complexity for matrix completion are removed by two-phase sampling; namely, assuming that the underlying matrix has a constant number $L$ of leverage scores within a factor of $\log^2(n)$ of either its largest row or column leverage score.

\begin{cor}
\label{cor:second}
\revise{Suppose the rank-$r$ matrix $M$ is such that $ \max\{ \mu_{(1)}, \nu_{(1)} \} \leq \frac{n}{r \log^2(n)}$ and such that
$$ \mu_{(L)} \leq \mu_{(1)}/\log^2(n), \quad  \nu_{(L)} \leq \nu_{(1)}/\log^2(n)$$
for some fixed integer $L$. Assuming also the mild condition that~\eqref{eq:n4lowerbound} holds for all $i,j \in [1:n]$, the two-phase algorithm $\alg$ with sampling probability $$
p = C_3 \cdot \frac{r}{n} \cdot \max\{ \frac{L \tau^{-1}  \kappa^4}{\log^2(n)}, 1 \} \left( (L+1) (\nu_{(1)}+  \mu_{(1)}) \right)
$$
recovers $M$ as $\widehat{M} = M$ with probability exceeding $1-(\tau + \frac{1}{n^{10}})$ from an expected total number of samples
\begin{align}
\label{power:complexity}
\mathbb{E} | \Omega | &\leq 2 p n^2 + 6C_2 r n \kappa^2 \log^2(n) \lesssim r n (L+1)(\nu_{(1)}+  \mu_{(1)}) \max\{ \frac{L \tau^{-1}  \kappa^4}{\log^2(n)}, 1 \}  + r n \kappa^2 \log^2(n).
\end{align}
Above, $C_2$ is the universal constant from Proposition \ref{thm:gen Rachel's thm}, and $C_3$ is the universal constant from Theorem~\ref{thm:main}.}
\end{cor}

\revise{In the sample complexity appearing in Corollary~\ref{cor:second}, the term $r n (L+1)(\nu_{(1)}+  \mu_{(1)}) \max\{ \frac{L \tau^{-1}  \kappa^4}{\log^2(n)}, 1 \}$ which relates to the matrix coherence $\eta(M)$ does not involve a multiplicative factor of $\log^2(n)$, unlike the guarantee for UMC.}

\begin{subsection}{Proofs}
\label{sec:proofs}

\revise{This section contains proofs of our results.}

\begin{subsubsection}{Proof of Lemma~\ref{lemma:estLev}}

\begin{proof}
Without loss of generality, assume that the rows and columns of $M$ are arranged according to decreasing magnitude of their corresponding leverage scores, that is, $\mu_1 \geq \mu_2 \geq \dots \geq \mu_n$ and $\nu_1 \geq \nu_2 \geq \dots \geq \nu_n$.

We make use of several inequalities relating leverage scores, singular values, and magnitudes of matrix entries.  Noting that
$M[i,j] =  \sum_{k=1}^r U[i,k] \sigma_k V[j,k],$ we have
$$
| M[i,j] | \leq \frac{r}{n} \sigma_1 \sqrt{ \mu_i \nu_j}
$$
and
\begin{align}
 \frac{r}{n} \sigma_r^2 \mu_i &\leq \sum_{j=1}^n | M[i,j] |^2 \leq \frac{r}{n} \sigma_1^2 \mu_i, \nonumber \\
  \frac{r}{n} \sigma_r^2 \nu_j &\leq \sum_{i=1}^n | M[i,j] |^2 \leq \frac{r}{n} \sigma_1^2 \nu_j, \nonumber \\
    r \sigma_r^2  &\leq \sum_{i,j=1}^n | M[i,j] |^2 \leq r \sigma_1^2.
 \end{align}
We now apply Bernstein's and Chebyshev's concentration inequalities using these bounds:
\begin{enumerate}
\item First, consider $X_{i,j} = \frac{(Y[i,j])^2}{p} - | M[i,j] |^2$, $(i,j) \in [1:n]^2$.  Note that $X_{i,j}$ are independent zero-mean random variables.  Further, $|X_{i,j} | \leq \frac{r^2}{p n^2} \sigma_1^2 \eta^2$ and
$$\sum_{i,j=1}^n \mathbb{E}[X_{i,j}^2] \leq \frac{1}{p}  \sum_{i,j=1}^n | M[i,j]|^4 \leq \frac{1}{p} \frac{r^2}{n^2} \sigma_1^2 \eta^2 \| M \|_F^2.$$
By Bernstein's inequality and the above bounds,
$$
\text{Prob} \left( | \frac{1}{p} \| Y \|_F^2 - \| M \|_F^2 | > \frac{1}{2}\| M \|_F^2 \right) \leq 2\tau,
$$
provided that $p n^2 \geq 10 \kappa^2 n (\frac{r}{n} \eta^2) \log(1/\tau) \geq 10 \kappa^2 r \eta^2 \log(1/\tau)$.  In particular, in this event,
$$
\frac{1}{2} r \sigma_r^2 \leq \frac{1}{p} \| Y \|_F^2  \leq \frac{3}{2} r \sigma_1^2.
$$

\item Second, fix index $i$ and consider $X_{j} = \frac{(Y[i,j])^2}{p} - | M[i,j] |^2$, $j \in [1:n]$.  The $X_{j}$ are independent zero-mean random variables, and
\begin{align}
\sum_j \mathbb{E} X_j^2 &\leq \frac{1}{p} \sum_j | M[i,j] |^4 \nonumber \\
&\leq \frac{1}{p} \left( M[i,1]^4 + \dots + M[i,d_1]^4 + \left( \max_{d_1+1 \leq j \leq n} | M[i,j] |^2 \right) \revise{\sum_{j=1}^n | M[i,j] |^2} \right) \nonumber \\
&\leq \frac{1}{p}  \left( \frac{r^4}{n^4} \sigma_1^4 \mu_i^2  \sum_{j=1}^{d_1} \nu_j^2 +   \frac{r^3}{n^3} \sigma_1^4 \mu_i^2 \nu_{d_1+1} \right).
\end{align}
Applying Chebyshev's inequality,
$$
\text{Prob} \left( | \frac{1}{p} \| Y[i,:] \|_F^2 - \| M[i,:] \|_F^2 | > \frac{1}{2} \frac{r}{n} \sigma_r^2 \mu_i  \right) \leq \tau,
$$
provided that
$$p n^2 \geq \frac{4 \kappa^4 r n \left(( \frac{r}{n} \sum_{j=1}^{d_1} \nu_j^2 )  + \nu_{d_1+1} \right)}{\tau}.$$  In particular, in case of this event, $$
\frac{1}{2} \frac{r}{n} \sigma_r^2 \mu_i \leq \frac{1}{p} \| Y[i,:] \|_F^2 \leq \frac{3}{2} \frac{r}{n} \sigma_1^2 \mu_i.
$$
Apply the union bound to this inequality over $i \in \{1,2, \dots, L\}$, and repeat the process over columns $j \in \{1,2, \dots, L\}$.
\end{enumerate}
Taking the union bound over events 1 and 2, and noting that $1/\tau \geq \log(4/\tau)$ for $0 \leq \tau \leq 1/3$, we find that as long as the sampling probability $p \in (0,1]$ is sufficiently large that
$$
pn^2 \geq \frac{16L \kappa^4 r n  \left(( \frac{r}{n} \sum_{j=1}^{d_1} \nu_j^2 ) + \nu_{d+1} \right)}{\tau},
$$
it holds with probability at least $1-\tau$ that
\begin{align}
\frac{\mu_i}{3 \kappa(M)^2} &\leq \frac{n \| Y[i,:] \|_F^2}{ \| Y \|_F^2 } \leq 3 \kappa(M)^2 \mu_i, \quad i \in [1:L] \nonumber \\
\frac{\nu_j}{3 \kappa(M)^2}  &\leq \frac{n \| Y[:,j] \|_F^2}{ \| Y \|_F^2 } \leq 3 \kappa(M)^2 \nu_j, \quad j \in [1:L].  \nonumber
\end{align}

\end{proof}

\begin{subsubsection}{Proof of Theorem \ref{thm:main}}

In this section, we prove the main result, Theorem \ref{thm:main}.

\begin{proof}
 Without loss of generality, assume the rows and columns of $M$ are ordered according to the decreasing rearrangement of their leverage scores: $\mu_1 \geq \mu_2 \geq \dots \geq \mu_n$ and $\nu_1 \geq \nu_2 \geq \dots \geq \nu_n$.

 \revise{First, note that for any value of $p$ used in Algorithm $\alg$, the expected total number of sampled entries, $|\Omega|,$ after Phase 1 and Phase 2 of Algorithm $\alg$ is bounded by
 \begin{align}
 \mathbb{E} | \Omega | & \leq 2 p n^2 + \frac{3C_2 r \log^2(n)}{n}  \left[ \sum_{ i, j \in [1:n] } (\widehat{\mu_i} + \widehat{\nu_j})  \right]  \nonumber \\
  &=  2 p n^2 + 6C_2 r n \kappa^2 \log^2(n),  \nonumber
 \end{align}
where $C_2$ is the universal constant from Proposition \ref{thm:gen Rachel's thm} and the second line follows from~\eqref{eq:mu bar alg}.}

\revise{Second, note that the condition \eqref{rec:newnew} need only be satisfied for one value of $L \in [0:n]$. If \eqref{rec:newnew} is satisfied with $L = 0$, then
\[
p \geq \max\{16,C_1,C_2\} \cdot \frac{r}{n} \cdot \max\{ 0 \tau^{-1}  \kappa^4, \log^2(n) \} \left(  \nu_{1} + \mu_{1} \right) \\
\geq C_1 \frac{r}{n} \log^2(n) \max\{ \nu_{1}, \mu_{1}\},
\]
which according to the classical Proposition~\ref{prop:vanilla MC} is a sufficient condition for exact matrix recovery (with probability at least $1-\frac{1}{n^{10}}$) when entries are sampled uniformly with probability $p$. In Algorithm $\alg$, the Phase 1 entries are sampled uniformly with probability $p$ and constitute a subset of the total sample set $\Omega$ used for completion after Phase 2. A sufficient condition for exact recovery on this subset is immediately a sufficient condition for exact recovery using the larger set of measurements $\Omega$. Therefore, if \eqref{rec:newnew} is satisfied with $L = 0$, we conclude that $\widehat{M} = M$ with probability exceeding $1 - \frac{1}{n^{10}}$.}

\revise{Finally, if \eqref{rec:newnew} is satisfied for some $L > 0$, note that with this value of $L$,
\begin{align*}
pn^2 & \geq n^2 \max\{16,C_1,C_2\} \cdot \frac{r}{n} \max\{ L \tau^{-1}  \kappa^4, \log^2(n) \} \left(  \frac{r}{n} \sum_{j=1}^{L} \nu_{j}^2 + \nu_{L+1} +  \frac{r}{n} \sum_{i=1}^{L} \mu_{i}^2   + \mu_{L+1} \right) \\
& \ge 16 rn L \tau^{-1}  \kappa^4 \left(  \frac{r}{n} \sum_{j=1}^{L} \nu_{j}^2 + \nu_{L+1} +  \frac{r}{n} \sum_{i=1}^{L} \mu_{i}^2   + \mu_{L+1} \right),
\end{align*}
which according to Lemma \ref{lemma:estLev} is a sufficient condition for the top $L$ row- and column-leverage scores to be estimated accurately using the uniformly sampled entries from Phase 1. In particular, with probability at least $1-\tau$, the estimates satisfy
\begin{align}
\frac{\mu_i}{3} &\leq  \widehat{\mu}_i, \quad \quad i \in [1:L] \nonumber \\
\frac{\nu_j}{3}  &\leq \widehat{\nu}_j, \quad \quad j \in [1:L].  \nonumber
\end{align}
In this event, it follows from
\[
P[i,j] = \min\{1, 3\revise{C_2} \frac{r \log^2(n)}{n}(\widehat{\mu}_i + \widehat{\nu}_j) \}
\]
that for the top row- and column-leverage scores,
\begin{align*}
P[i,j] & \geq \min\{1, \revise{C_2} \frac{r \log^2(n)}{n}(\mu_i + \nu_j) \}, \quad \quad i,j \in [1:L], \\
P[i,j] & \geq \min\{1, \revise{C_2} \frac{r \log^2(n)}{n}\mu_i \}, \quad \quad i \in [1:L], j \notin [1:L], \\
P[i,j] & \geq \min\{1, \revise{C_2} \frac{r \log^2(n)}{n} \nu_j \}, \quad \quad j \in [1:L], i \notin [1:L].
\end{align*}
Moreover, for the remaining row- and column-leverage scores, \eqref{rec:newnew} ensures that
\begin{align*}
p & \geq \max\{16,C_1,C_2\} \cdot \frac{r}{n} \max\{ L \tau^{-1}  \kappa^4, \log^2(n) \} \left(  \frac{r}{n} \sum_{j=1}^{L} \nu_{j}^2 + \nu_{L+1} +  \frac{r}{n} \sum_{i=1}^{L} \mu_{i}^2   + \mu_{L+1} \right) \\
& \ge C_2 \frac{r}{n} \log^2(n) \left(  \nu_{L+1} +  \mu_{L+1} \right),
\end{align*}
from which it follows that
\begin{align*}
p &\geq \revise{C_2}(\mu_{i} + \nu_{j}) \frac{r \log^2(n)}{n}, \quad \quad i,j \in [L+1: n], \\
p &\geq \revise{C_2}\mu_{i} \frac{r \log^2(n)}{n}, \quad \quad i \in [L+1: n], j \notin [L+1:n] \\
p &\geq \revise{C_2}\nu_{j} \frac{r \log^2(n)}{n}, \quad \quad j \in [L+1: n], i \notin [L+1:n].
\end{align*}
Thus in this event, consider the subset of the entries in $\Omega$ that are sampled fresh in Phase 2, i.e., where each entry $[i,j]$ is independently observed with probability $p + P[i,j]$, and
$$
p + P[i,j]  \geq \min\{1, \revise{C_2} \frac{r \log^2(n)}{n} (\mu_i + \nu_j) \}, \quad \quad \text{ for all} ~ i, j \in [1:n].
$$
A guarantee of exact recovery of $M$ then follows by applying the known result for leveraged matrix completion, Proposition \ref{thm:gen Rachel's thm}. In particular, this provides a guarantee for exact recovery (with probability at least $1-\frac{1}{n^{10}}$) using only the Phase 2 samples in $\Omega$. In the event that the matrix $M$ can be correctly recovered from the Phase 2 samples, however, adding in the Phase 1 samples to the index set $\Omega$ will only shrink the feasible set of the Program~\eqref{eq:pr Alg I}. Since the feasible set will always contain $M$, the exact recovery guarantee will also apply to the full set $\Omega$ containing both the Phase 1 and Phase 2 samples. Therefore, if \eqref{rec:newnew} is satisfied for some $L > 0$, we conclude that $\widehat{M} = M$ with probability exceeding $1 - (\tau+\frac{1}{n^{10}})$.
}
\end{proof}

\begin{subsubsection}{Proof of Corollary~\ref{cor:main}}

\begin{proof}
\revise{Without loss of generality, suppose $\eta = \mu_{(1)} \geq \nu_{(1)}$. Set $L = \lceil{ \eta^{\frac{1}{1+T}} \rceil}$ and note that
\begin{align}
p &= 8C_3 \tau^{-1} \kappa^4 \frac{r^2}{n^2} \eta^{\frac{3+2T}{1+T}} \nonumber \\
&\geq 4C_3 \eta^{\frac{1}{1+T}} \tau^{-1} \kappa^4 \frac{r}{n} \left( \frac{r}{n} \eta^2 +  1 \right)  \quad \quad \quad \quad \quad \quad \quad \quad \quad \quad \quad \textit{using $\eta^2 \geq \frac{n}{r}$}  \nonumber \\
&\geq 2C_3 L \tau^{-1} \kappa^4 \frac{r}{n} \left( \frac{r}{n} \eta^2 +  \eta (L+1)^{-(1+T)} \right)  \quad \quad \quad \quad \quad \quad \quad  \textit{setting $L = \lceil{ \eta^{\frac{1}{1+T}} \rceil}$}  \nonumber \\
&\geq 2C_3 L \tau^{-1} \kappa^4 \frac{r}{n} \left( \frac{r}{n} \eta^2 \sum_{i=1}^L   i^{-2(1+T)} + \eta (L+1)^{-(1+T)} \right)  \nonumber \\
&\geq 2C_3 L \tau^{-1} \kappa^4 \frac{r}{n} \left( \frac{r}{n} \sum_{i=1}^L \mu_{(i)}^2 + \mu_{(L+1)} \right) \nonumber \\
&\geq C_3 L \tau^{-1} \kappa^4 \frac{r}{n} \left( \frac{r}{n} \sum_{i=1}^L \mu_{(i)}^2 + \nu_{(L+1)} +  \frac{r}{n} \sum_{i=1}^L \nu_{(i)}^2 + \nu_{(L+1)}  \right) \\
&\geq C_3 \cdot \frac{r}{n} \cdot \max\{ L \tau^{-1}  \kappa^4, \log^2(n) \} \left(  \frac{r}{n} \sum_{j=1}^{L} \nu_{(j)}^2 + \nu_{(L+1)} +  \frac{r}{n} \sum_{i=1}^{L} \mu_{(i)}^2 + \mu_{(L+1)} \right),
\end{align}
where the last line follows because $L \tau^{-1} \kappa^4 \geq L \geq \log^2(n)$. Thus, with this choice of $p$, we see that~\eqref{rec:newnew} is satisfied with $L = \lceil{ \eta^{\frac{1}{1+T}} \rceil}$. Theorem~\ref{thm:main} then gives the corollary.}
\end{proof}

\begin{subsubsection}{Proof of Corollary~\ref{cor:second}}

\begin{proof}
\revise{We have
\begin{align*}
   p &= C_3 \cdot \frac{r}{n} \cdot \max\{ \frac{L \tau^{-1}  \kappa^4}{\log^2(n)}, 1 \} \left( (L+1) (\nu_{(1)}+  \mu_{(1)}) \right) \\
   & \ge C_3 \cdot \frac{r}{n} \cdot \max\{ L \tau^{-1}  \kappa^4, \log^2(n) \} \left( (L/\log^2(n)) \nu_{(1)} + \nu_{(L+1)} + (L/\log^2(n)) \mu_{(1)}   + \mu_{(L+1)} \right) \\
   & \ge C_3 \cdot \frac{r}{n} \cdot \max\{ L \tau^{-1}  \kappa^4, \log^2(n) \} \left(  \frac{r}{n} L \nu_{(1)}^2 + \nu_{(L+1)} +  \frac{r}{n} L \mu_{(1)}^2   + \mu_{(L+1)} \right) \\
   & \ge C_3 \cdot \frac{r}{n} \cdot  \max\{ L \tau^{-1}  \kappa^4, \log^2(n) \} \left(  \frac{r}{n} \sum_{j=1}^{L} \nu_{(j)}^2 + \nu_{(L+1)} +  \frac{r}{n} \sum_{i=1}^{L} \mu_{(i)}^2   + \mu_{(L+1)} \right).
\end{align*}
Thus, we see that~\eqref{rec:newnew} is satisfied with this value of $L$.}
\end{proof}

\section{Numerical Simulations \label{sec:Numerical-Simulations}}

\subsection{Setup}
\label{sec:expsetup}

In this section, we compare $\alg$ with UMC. We test these algorithms on $12$ different matrices, each of size $n \times n$ with $n = 100$ and having rank $r = 5$. These matrices are constructed to have a range of condition numbers $\kappa$ and coherence levels $\eta$.

Eight of these matrices, which we denote as P1--P8 and refer to as {\em power-law matrices}, are generated by setting $M' = D U \Sigma V^{*} D$, where $U$ and $V$ are generic $n \times n$ random matrices with orthonormal columns, $\Sigma$ is an $n \times n$ diagonal matrix with ones on the diagonal (for well-conditioned matrices) or with entries linearly spaced between $1$ and $n$ (for poorly-conditioned matrices), and $D$ is an $n \times n$ diagonal matrix with entries that follow the power-law, $D[i,i]\propto i^{-\gamma}$. We set $\gamma = 0$ for incoherent matrices, and $\gamma$ up to $2$ for coherent matrices. We then construct the test matrix $M$ by adjusting the singular values of $M'$ to be all ones (for well-conditioned matrices) or to be linearly spaced between $1$ and $n$ (for poorly-conditioned matrices).

The remaining four test matrices, which we denote as B1--B4 and refer to as {\em block diagonal matrices}, are constructed as $M = \text{diag}(B_1, B_2, \dots, B_r)$, where each block $B_k$ is of size $b_k \times b_k$ and all of its entries are equal to the same value $v_k$. In this construction, the singular values of $M$ are given by $\{ b_k v_k \}_{k=1}^r$, and the (row or column) leverage scores of $M$ are given by $\left\{ \frac{n}{r} \cdot \frac{1}{b_k} \right\}_{k=1}^r$, with the value $\frac{n}{r} \cdot \frac{1}{b_k}$ occurring with multiplicity $b_k$. Thus, the condition number of $M$ is given by
\[
\kappa = \frac{\max_k b_k v_k}{\min_j b_j v_j},
\]
and the coherence of $M$ is given by
\[
\eta = \frac{n}{r} \cdot \frac{1}{\min_k b_k}.
\]
We consider four distinct cases:
\begin{itemize}
\item Matrix B1 (incoherent, well-conditioned): We set $b_k = \frac{n}{r}$ for all $k$ to make $\eta$ as small as possible ($\eta = 1$). We then set $v_k = \frac{1}{b_k}$ to make all singular values the same, and thus $\kappa = 1$.
\item Matrix B2 (coherent, well-conditioned): We set $b_1 = 2$ and we set the rest of the $b_k$'s to be larger than $2$, in order to make $\eta$ nearly as large as possible ($\eta = \frac{n}{2r}$). We then set $v_k = \frac{1}{b_k}$ to make all singular values the same, and thus $\kappa = 1$.
\item Matrix B3 (incoherent, poorly-conditioned): We set $b_k = \frac{n}{r}$ for all $k$ to make $\eta$ as small as possible ($\eta = 1$). We then set $v_k = 1 + \frac{(n-1)(k-1)}{r-1}$ (linearly spaced between $1$ and $n$), to make the condition number $\kappa = n$.
\item Matrix B4 (coherent, poorly-conditioned): We set $b_1 = 2$ and we set the rest of the $b_k$'s to be larger than $2$, in order to make $\eta$ nearly as large as possible ($\eta = \frac{n}{2r}$). We then set $v_k = \frac{1}{b_k} \left(1 + \frac{(n-1)(k-1)}{r-1}\right)$, to make the singular values of $M$ increase in a linear sequence from $1$ to $n$, and thus the condition number is $\kappa = n$.
\end{itemize}

After the constructions above, all matrices are normalized such that $\|M\|_F = 1$. The 12 test matrices are displayed in Figure~\ref{fig:testmatrices}. Each matrix is displayed along with its condition number $\kappa$ and coherence $\eta$.

\begin{figure}[p]
\begin{center}
\includegraphics[width=1.8in]{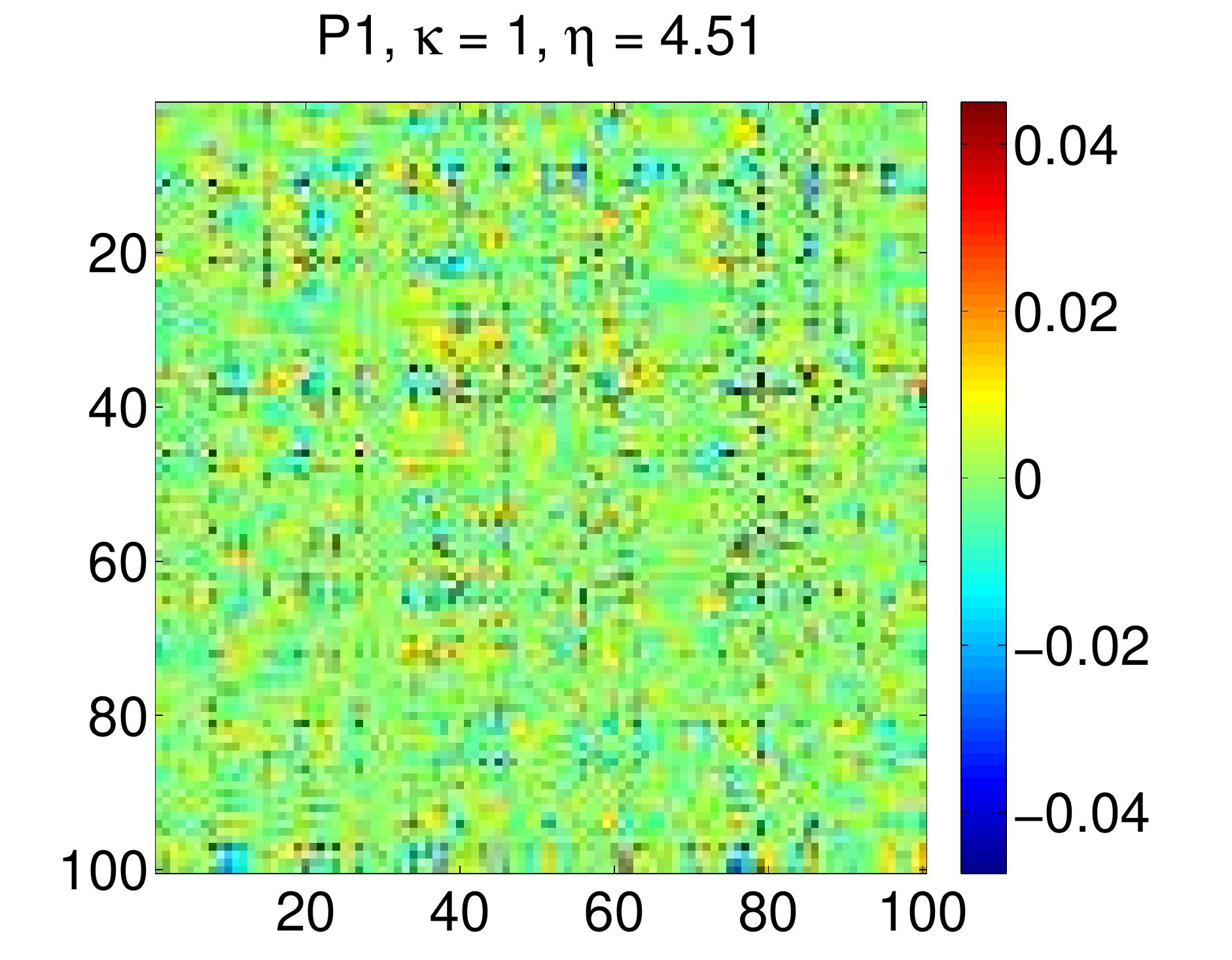}
\includegraphics[width=1.8in]{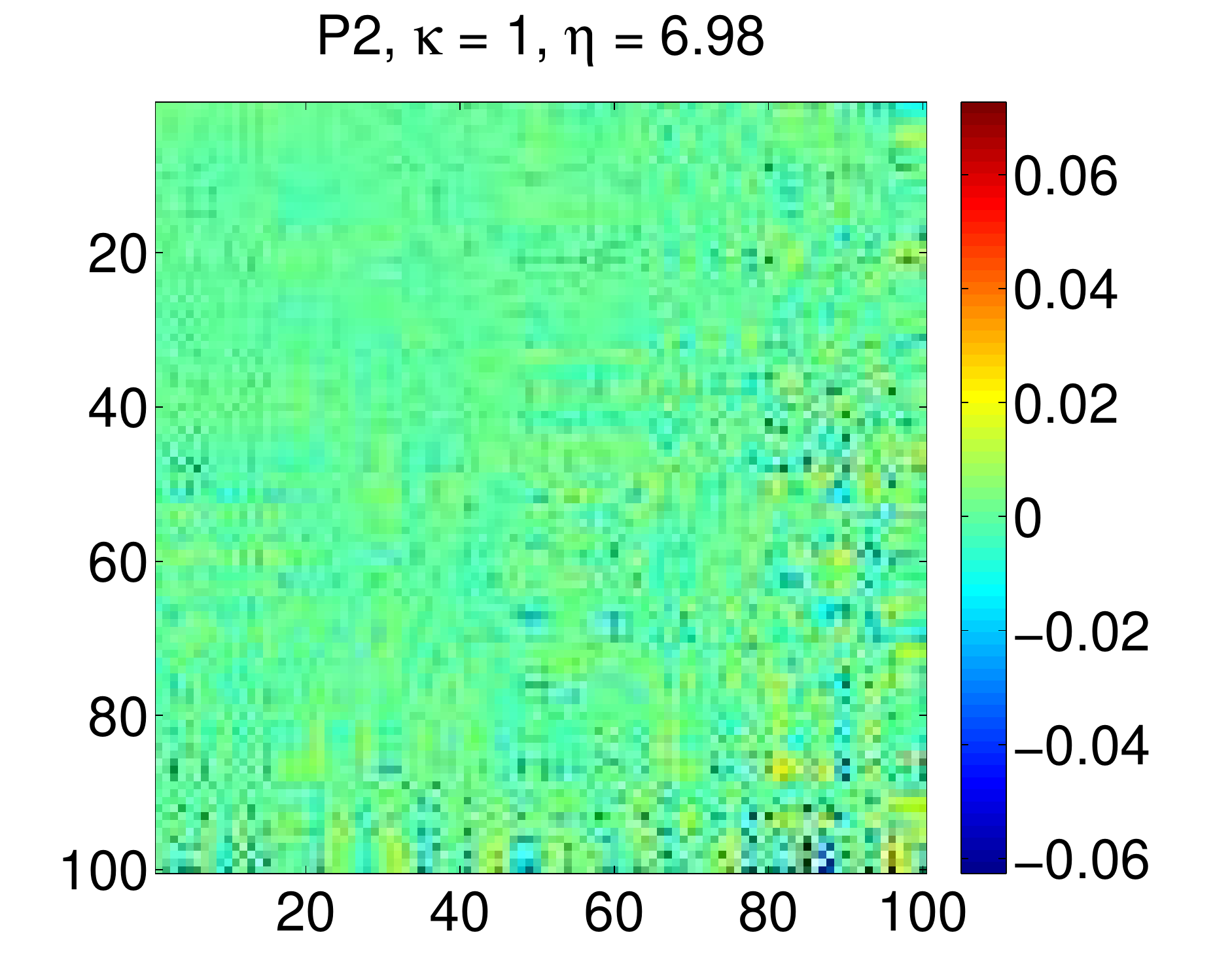}
\includegraphics[width=1.8in]{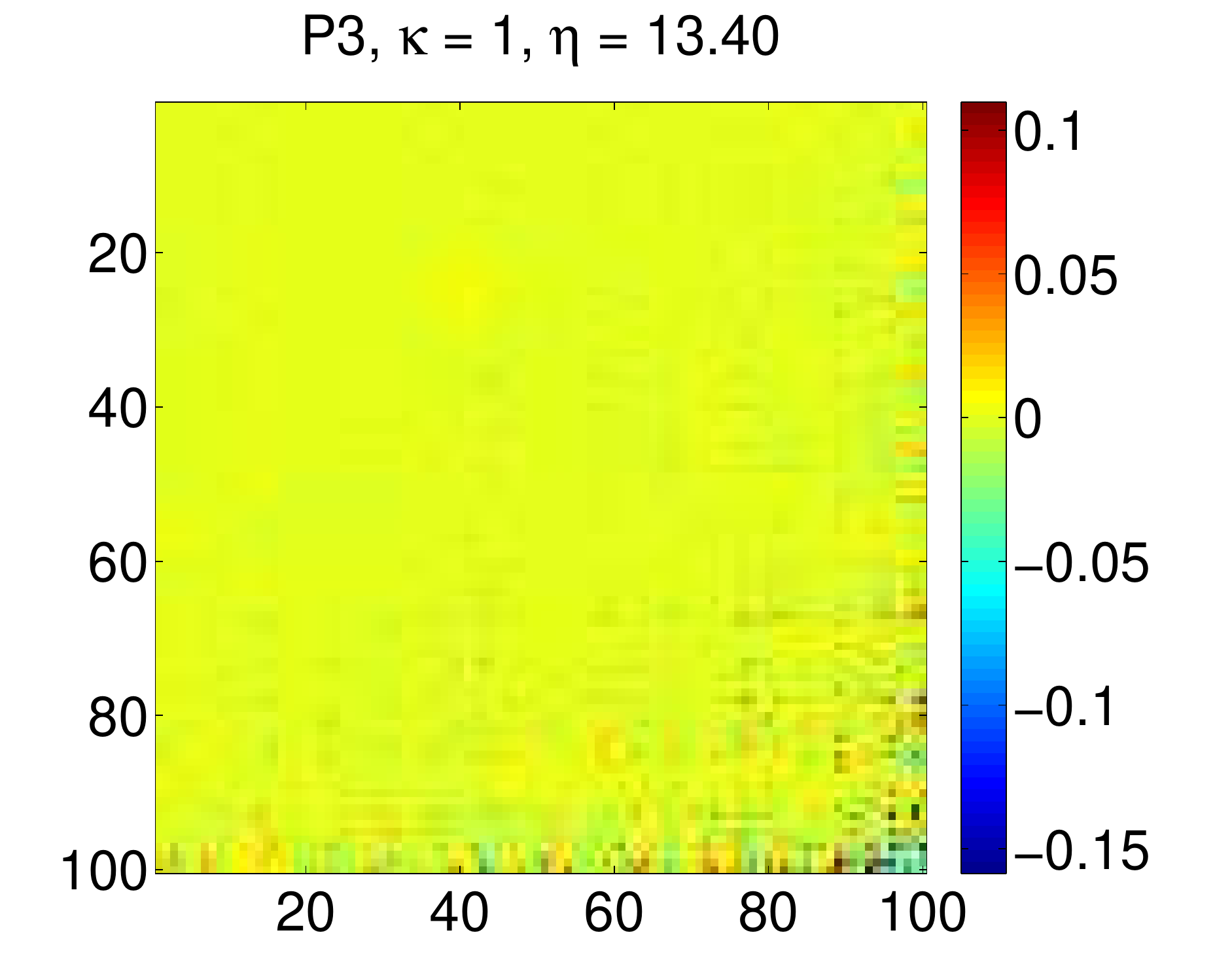}
\\
\includegraphics[width=1.8in]{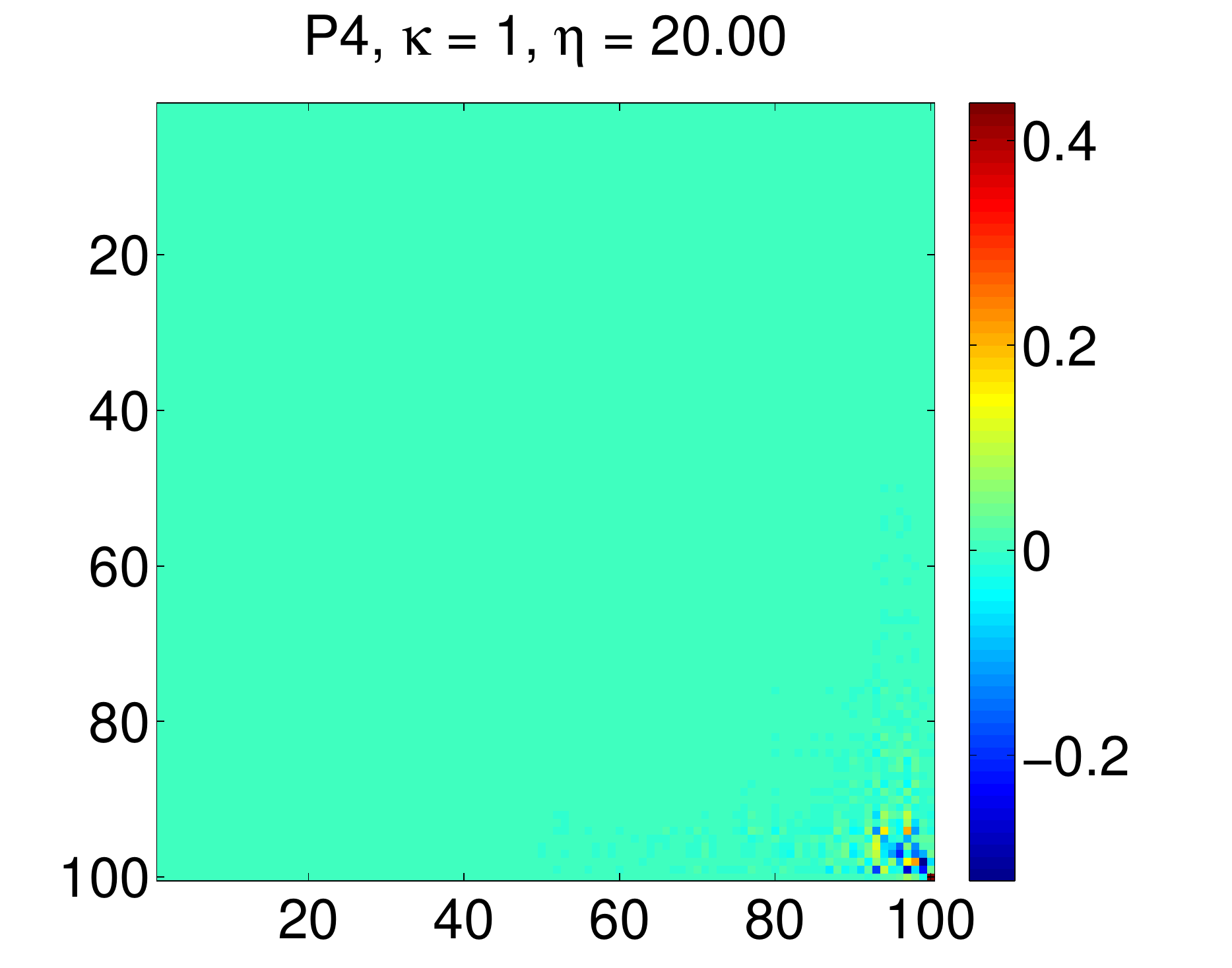}
\includegraphics[width=1.8in]{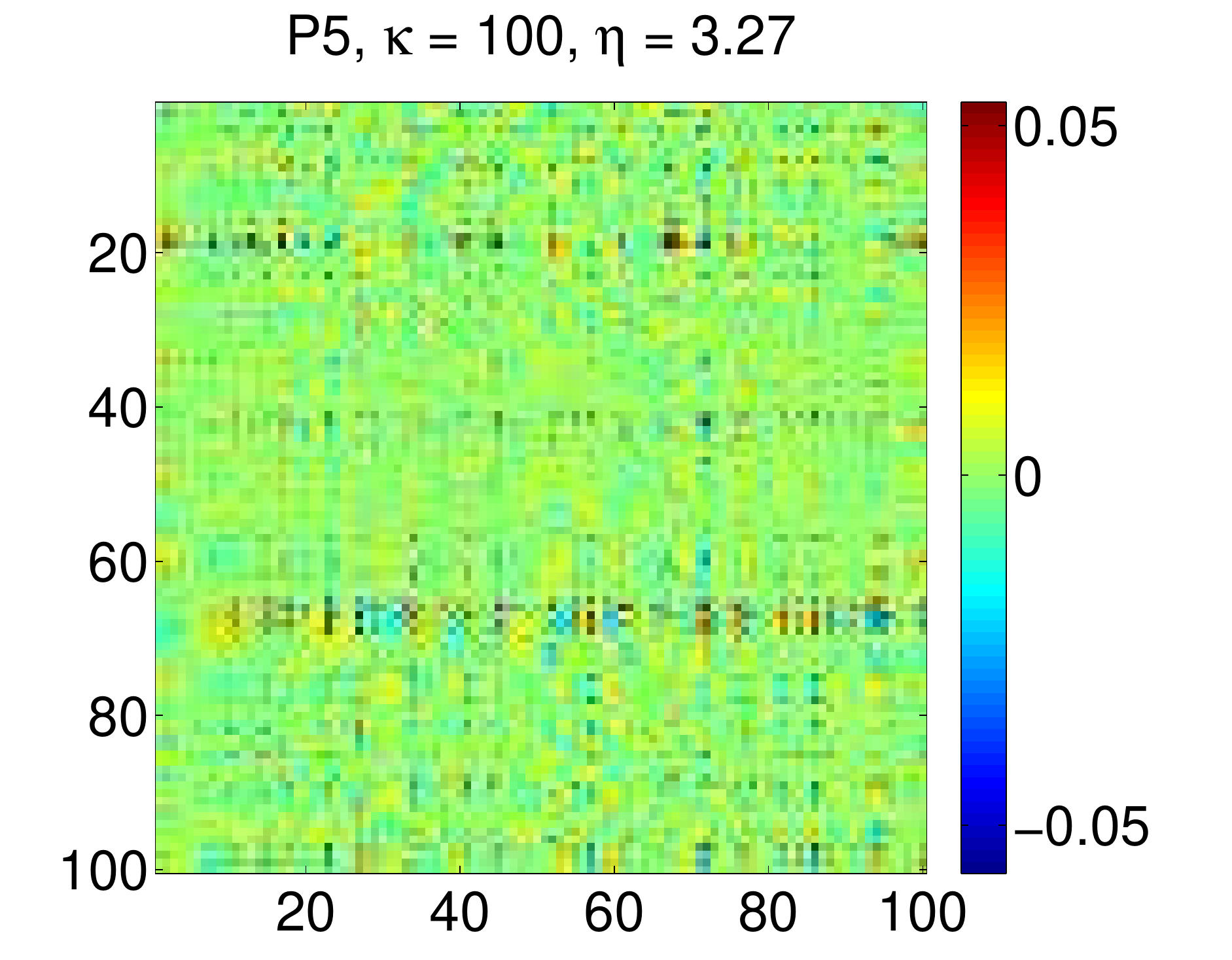}
\includegraphics[width=1.8in]{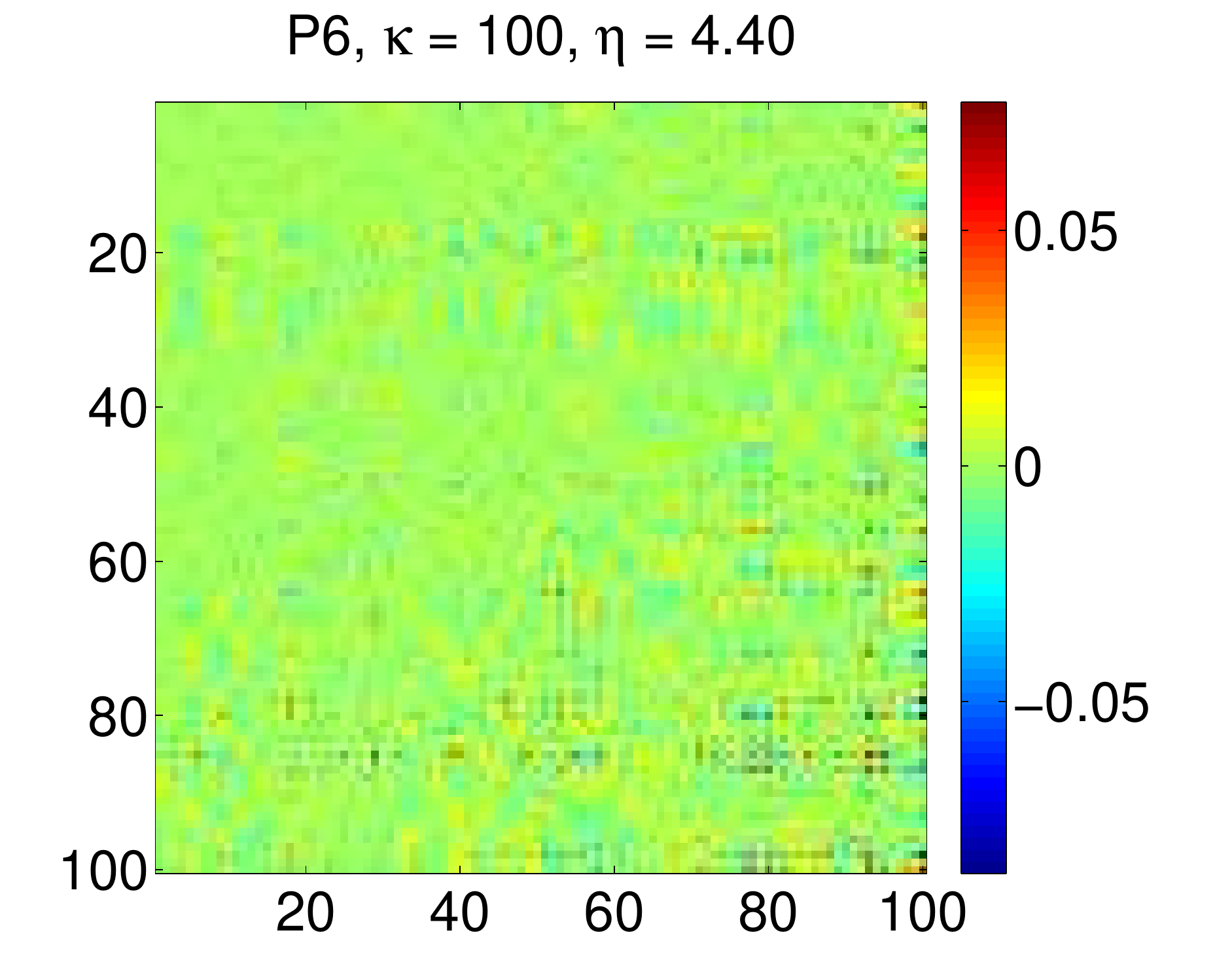}
\\
\includegraphics[width=1.8in]{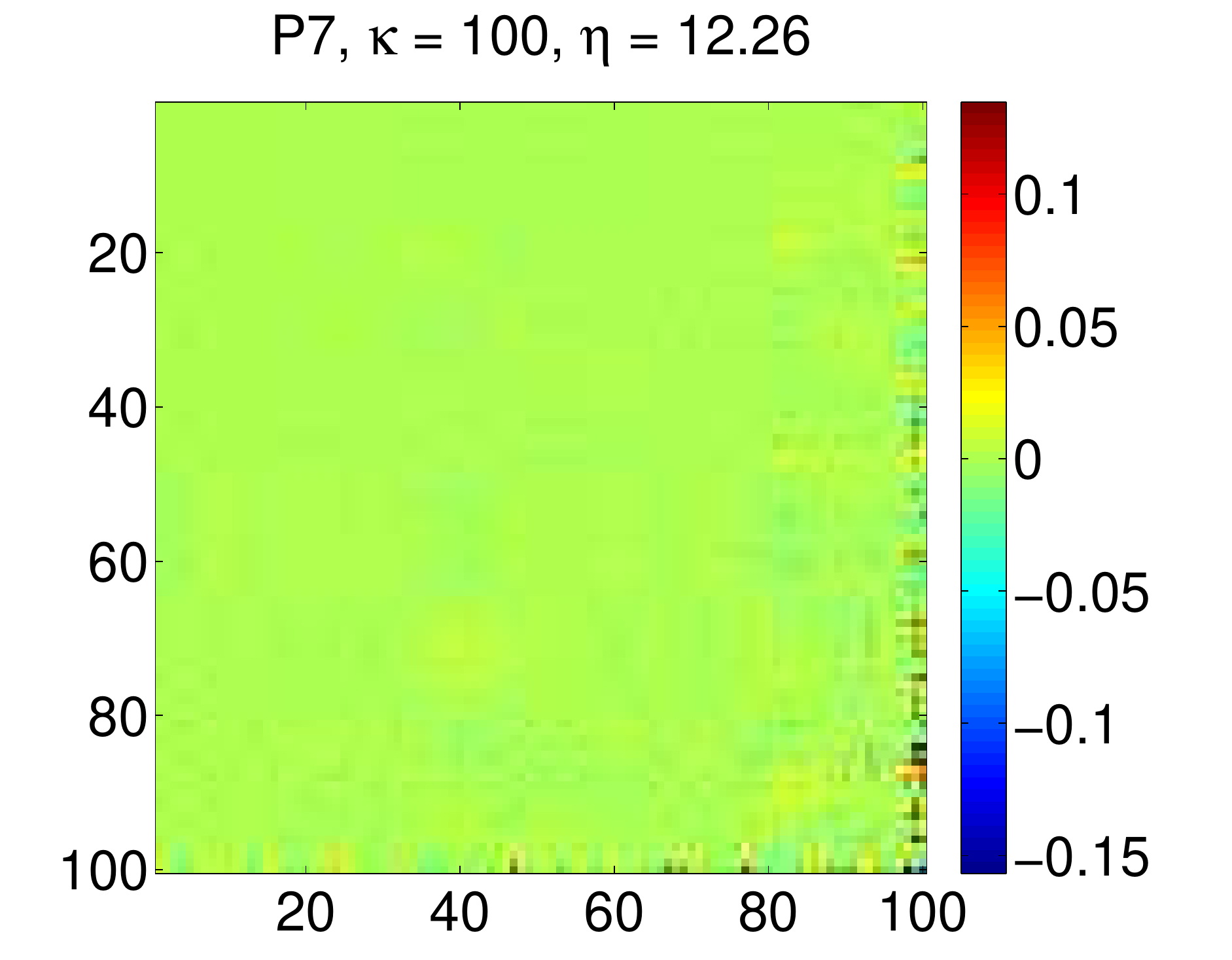}
\includegraphics[width=1.8in]{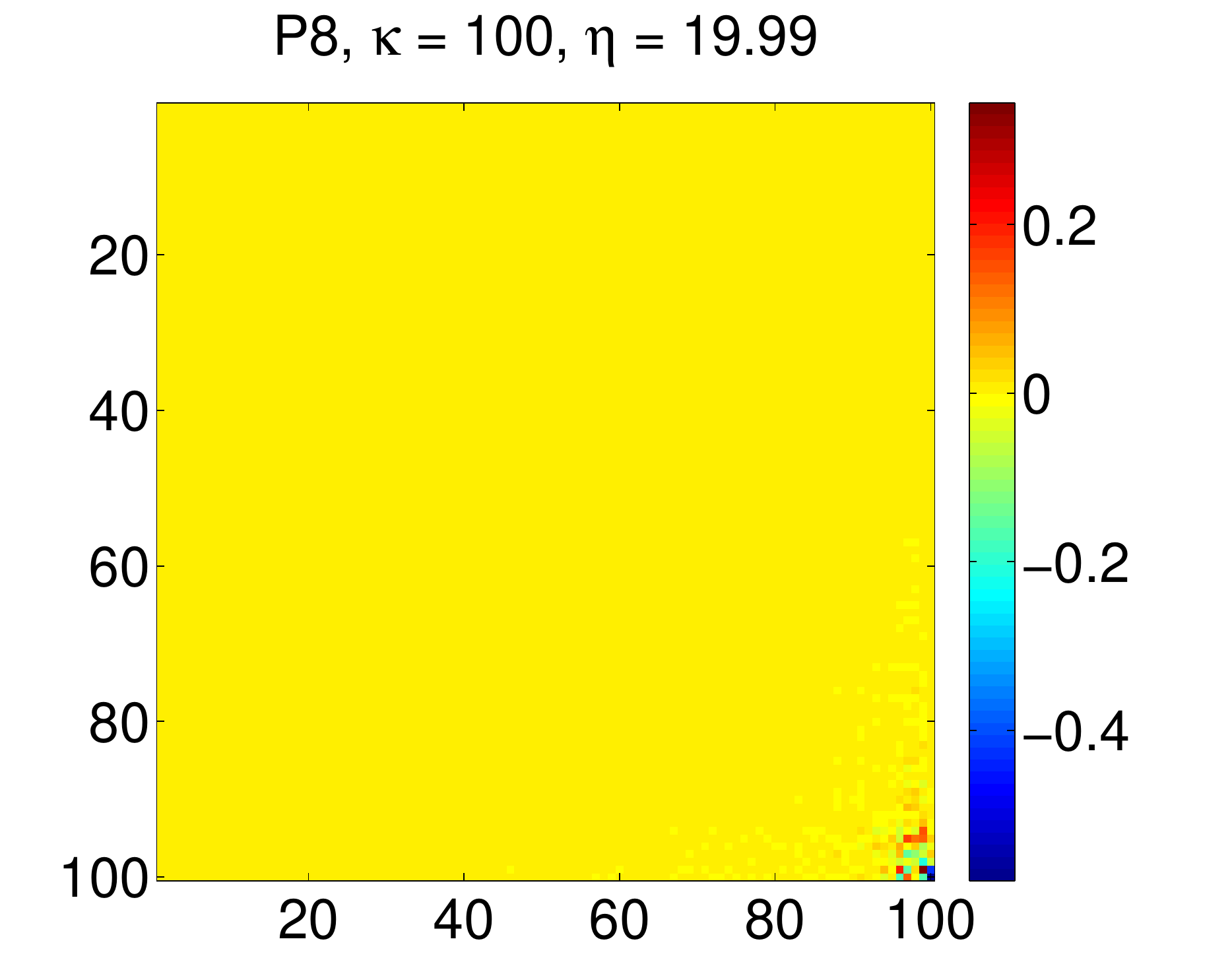}
\includegraphics[width=1.8in]{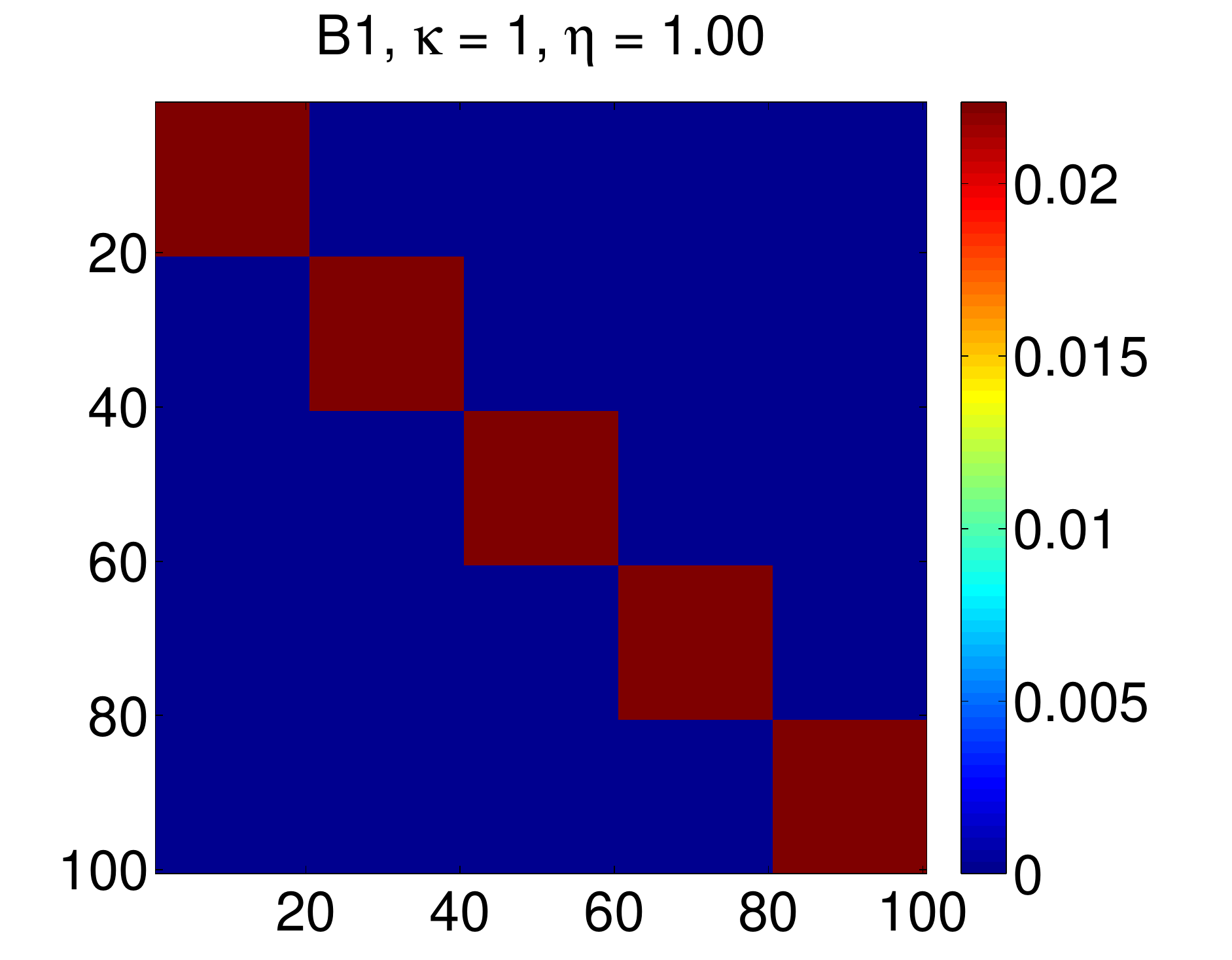}
\\
\includegraphics[width=1.8in]{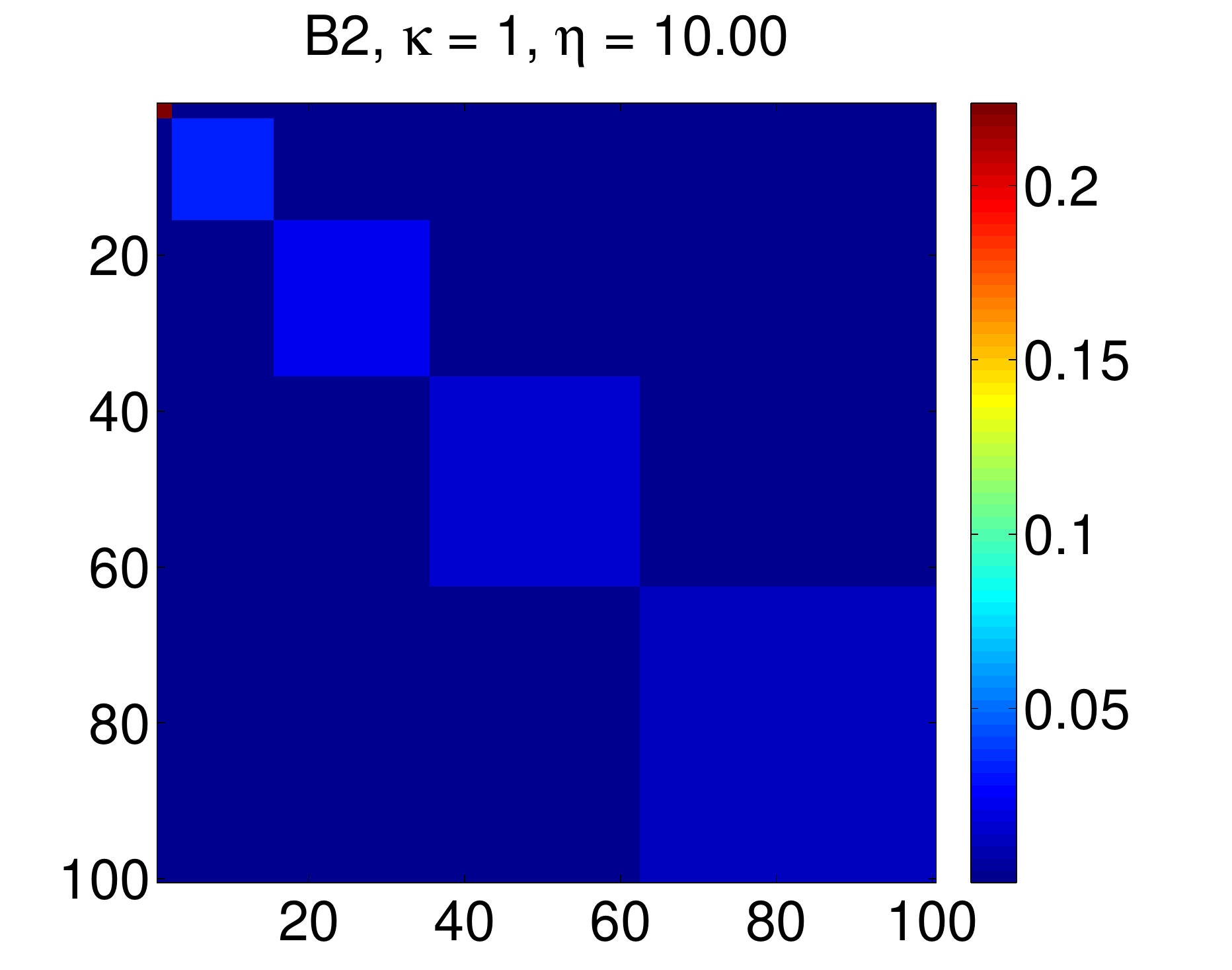}
\includegraphics[width=1.8in]{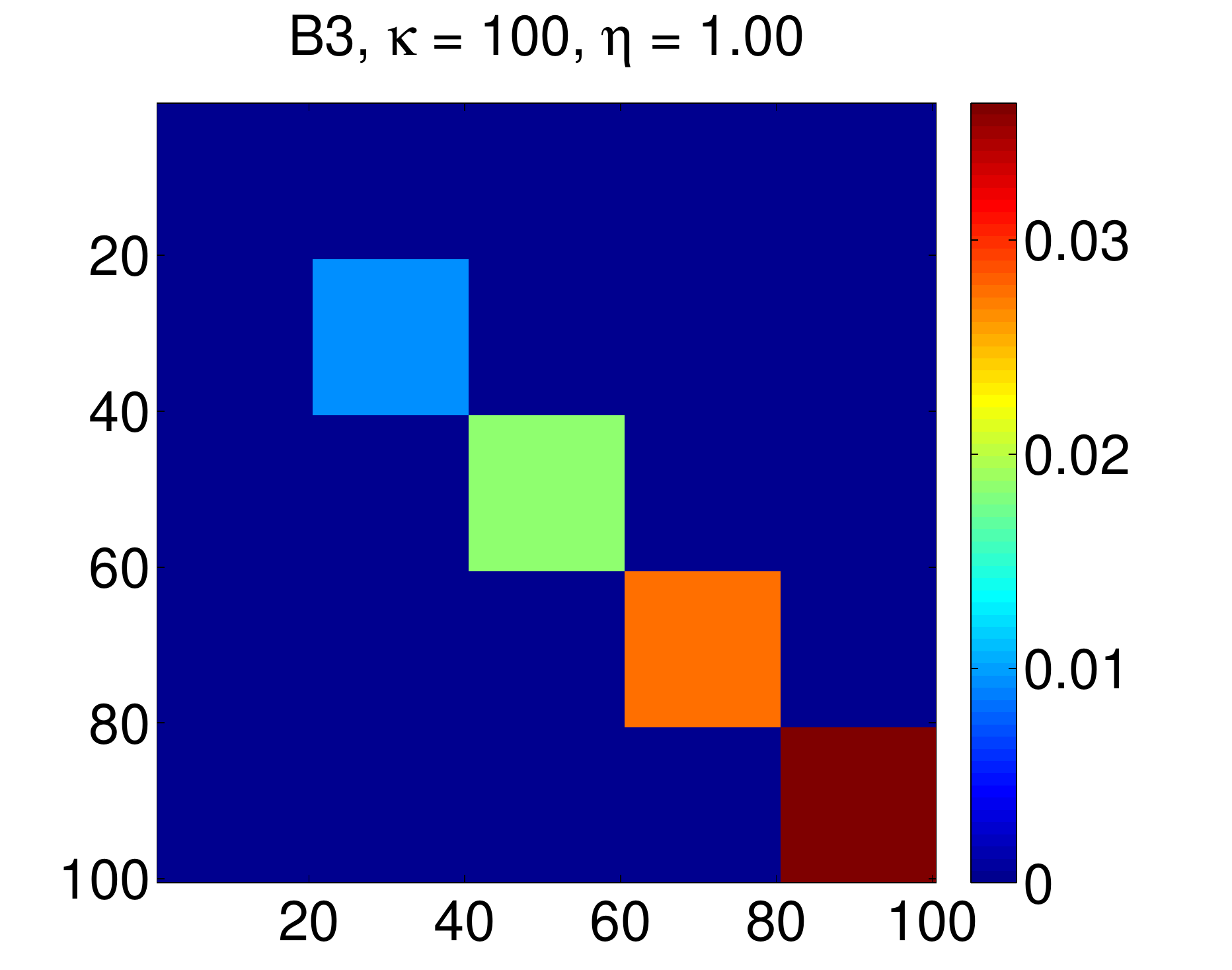}
\includegraphics[width=1.8in]{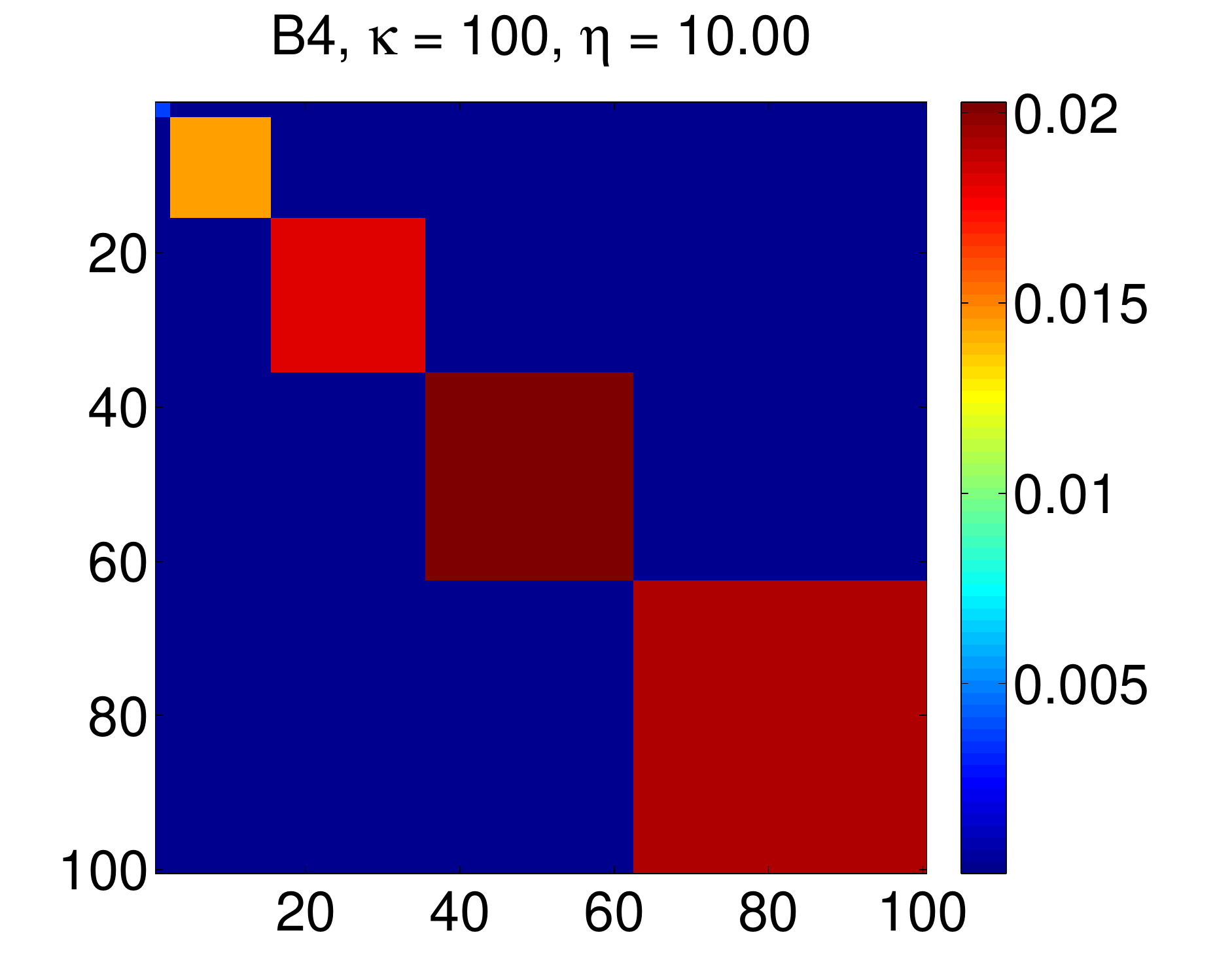}
\end{center}
\caption{\label{fig:testmatrices} Power-law matrices P1--P8 and block diagonal matrices B1--B4 used in experiments. }
\end{figure}

\revise{To provide a fair basis for comparing $\alg$ with UMC, we equalize the average sampling probability used in each algorithm. We refer to this average sampling probability as $q$ throughout our experiments. For UMC, to achieve an average sampling probability of $q$, we simply set $p = q$ and sample each entry of $M$ independently with probability $p$, as prescribed in Proposition~\ref{prop:vanilla MC}. For $\alg$, we achieve an average sampling probability of $q$ as follows:
\begin{itemize}
\item We first set $p = q/2$ and perform Phase 1 (uniform sampling with probability $p$) as prescribed in Figure~\ref{fig:MC:-A-two-phase}.
\item We then set Phase 2 sampling probabilities as
\begin{equation}
P[i,j] = \begin{cases} \min\{1, \frac{\beta r \log^2(n)}{n} \left(\widehat{\mu}_{i}+\widehat{\nu}_{j}\right)\}, & (i,j) \notin \Omega, \\ 0, & (i,j) \in \Omega, \end{cases}
\label{eq:def of Pij in alg modified}
\end{equation}
where $\beta$ is a constant chosen to achieve the desired average sampling probability, i.e., such that
\[
\sum_{i,j} P[i,j] = \frac{1}{2} q n^2,
\]
and we set $P[i,j] = 0$ for all pairs $(i,j) \in \Omega$ to avoid duplicate sampling locations with Phase 1. We sample nonuniformly according to the probabilities $P[i,j]$, and we add the sample indices to the Phase 1 index set $\Omega$ before finally completing the matrix.
\end{itemize}
We note that implementing $\alg$ in this fashion is a slight deviation from the formulation presented in Figure~\ref{fig:MC:-A-two-phase}, where for example $P[i,j]$ was set according to~\eqref{eq:def of Pij in alg}, and half of the uniform samples were reserved for Phase 2 only. While used in our proof, we do not believe that additional uniform samples are practically necessary for Phase 2, and so we omit them. Moreover, the implementation described above allows for careful testing of $\alg$ at a range of predetermined average sampling probabilities $q$. In this implementation, we note that~\eqref{eq:def of Pij in alg} and~\eqref{eq:def of Pij in alg modified} differ only in a scaling constant, and for sufficiently large values of $q$, ($i$) $p = q/2$ will exceed what is required by~\eqref{rec:newnew}, and ($ii$) $\beta$ will exceed $3 C_2$, and so $P[i,j]$ will exceed what is prescribed in~\eqref{eq:def of Pij in alg}. This implementation also reveals that, in practice, knowledge of the condition number $\kappa$ is not actually necessary: although $\kappa^2$ appears in the formulas for estimating the leverage scores in~\eqref{eq:mu bar alg}, this scaling is rendered irrelevant by our scaling with the constant $\beta$ in~\eqref{eq:def of Pij in alg modified} to hit our overall target sampling budget. Similarly, it is not actually necessary to know the rank $r$, which also appears in~\eqref{eq:def of Pij in alg modified}.
}

\revise{Finally, we note that we have experimented with different allocations of samples between Phase 1 and Phase 2 (setting $p = q/4$ and $p = 3q/4$ in Phase 1 and reserving the remainder of $q$ for Phase 2), and the results were not substantially different than those presented here. Thus, in practice, it appears reasonable to set $p = q/2$ in Phase 1.}

\subsection{Results on power-law matrices}
\label{sec:powerdiscussion}

We begin with tests on the well-conditioned power-law matrices P1--P4. Figure~\ref{fig:power}(a) shows the probability of exact recovery (declared with the relative reconstruction error $\| \widehat{M} - M \|_F / \|M\|_F < 10^{-4}$) as a function of the average sampling probability $q$, both for $\alg$ (solid curves) and for UMC (dashed curves). Results are averaged over 100 trials. In all cases, we see that $\alg$ performs better than UMC. We also see that the performance of both $\alg$ and UMC worsens as the coherence $\eta$ increases (moving from P1 toward P4). However, while the degradation in performance is quite dramatic for UMC, it is only minor with $\alg$. Figure~\ref{fig:power}(b) shows the average normalized recovery error $\| \widehat{M} - M \|_F / \|M\|_F$ over the 100 trials. Similar conclusions hold. For matrices P1 and P4 and with an average sampling probability of $q=0.3$, Figure~\ref{fig:power}(c) shows the accuracy of the leverage score estimates in $\alg$. In particular, for each row leverage score $\mu_i$ of $M$, we compute the average estimated leverage score $\widehat{\mu}_i$ and plot the relative multiplicative error $\widehat{\mu}_i/\mu_i$ versus the true leverage score $\mu_i$. Note that the range of the true leverage scores differs between matrix P1 (shown in blue) and matrix P4 (shown in red). However, we see that the relative multiplicative error of all estimates clusters around $1$ (dashed line), indicating that all leverage scores for these matrices can be estimated accurately. Finally, for matrices P1 and P4 and with an average sampling probability of $q=0.3$, Figure~\ref{fig:power}(d) shows the accuracy of the reconstructed columns of $M$. In particular, for each column $M[:,j]$, we plot the average relative squared reconstruction error $\|\widehat{M}[:,j]-M[:,j]\|_F^2/\|M[:,j]\|_F^2$ versus the true squared column norm $\|M[:,j]\|_F^2$, both for $\alg$ and for UMC. Both algorithms perform well on matrix P1; $\alg$ outperforms UMC on matrix P4; and the errors are relatively constant with respect to $\|M[:,j]\|_F^2$.

\begin{figure}[p]
\begin{center}
(a) \includegraphics[width=2.2in]{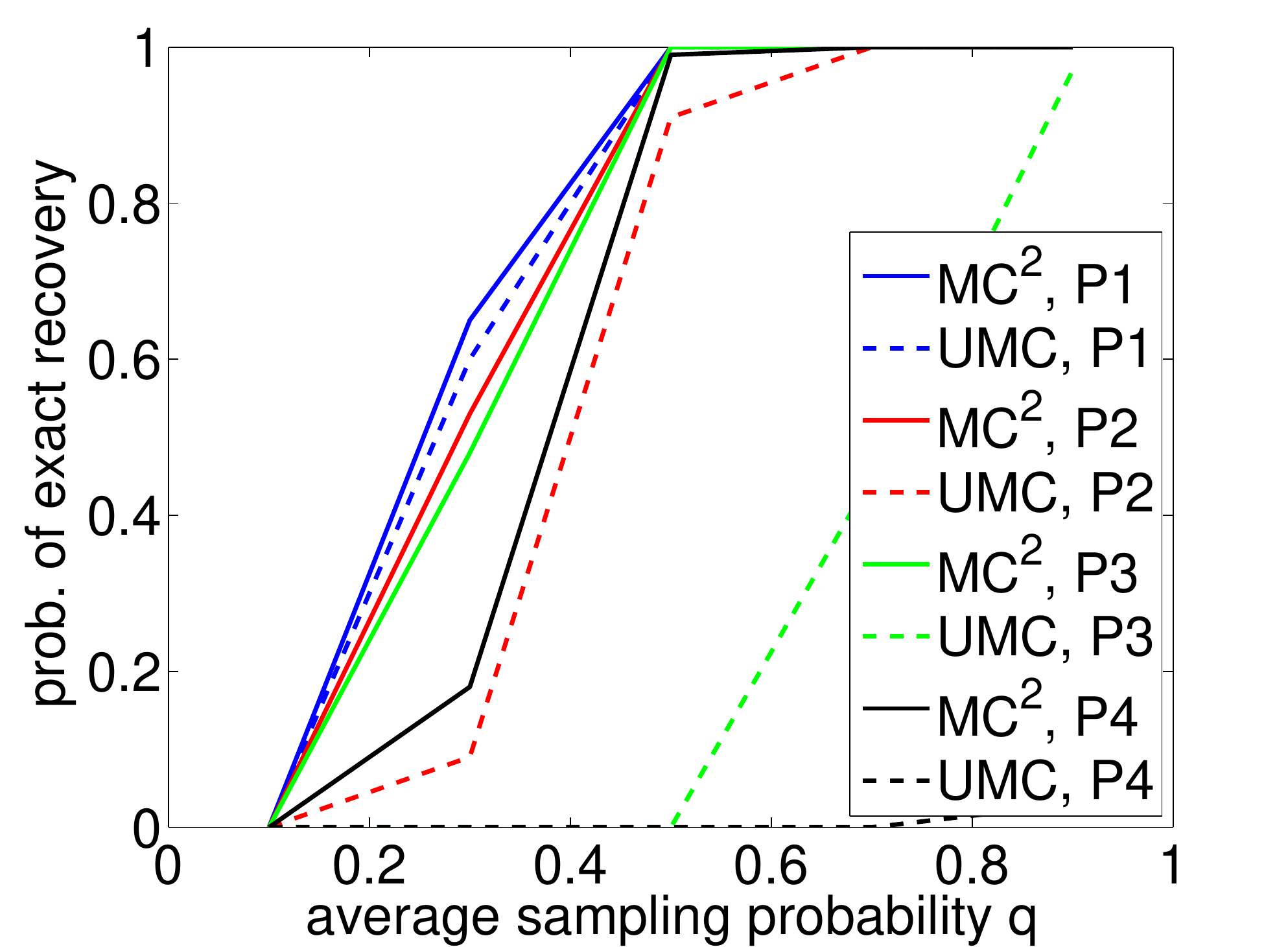} ~~~
(b) \includegraphics[width=2.2in]{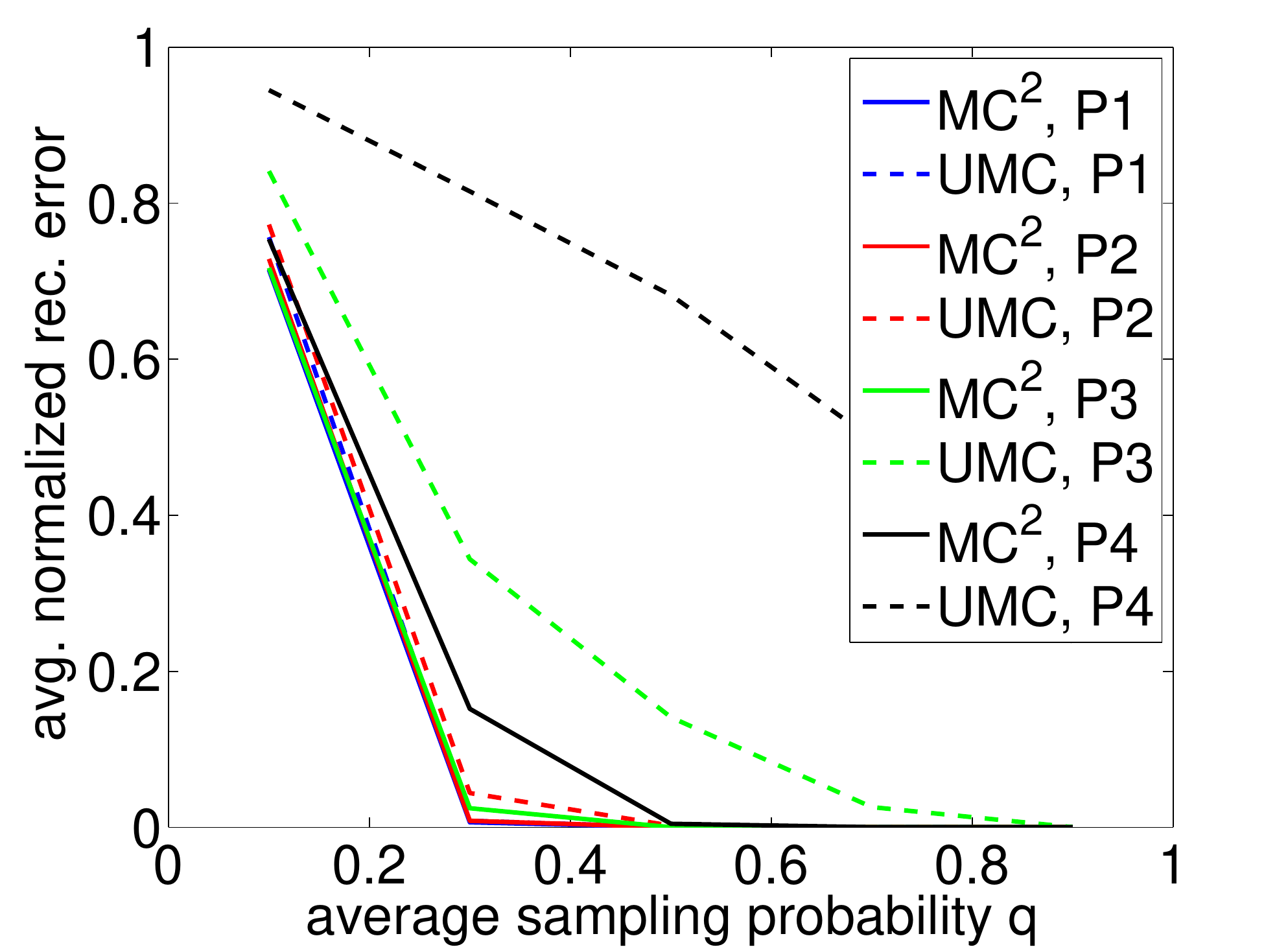}
\\
(c) \includegraphics[width=2.2in]{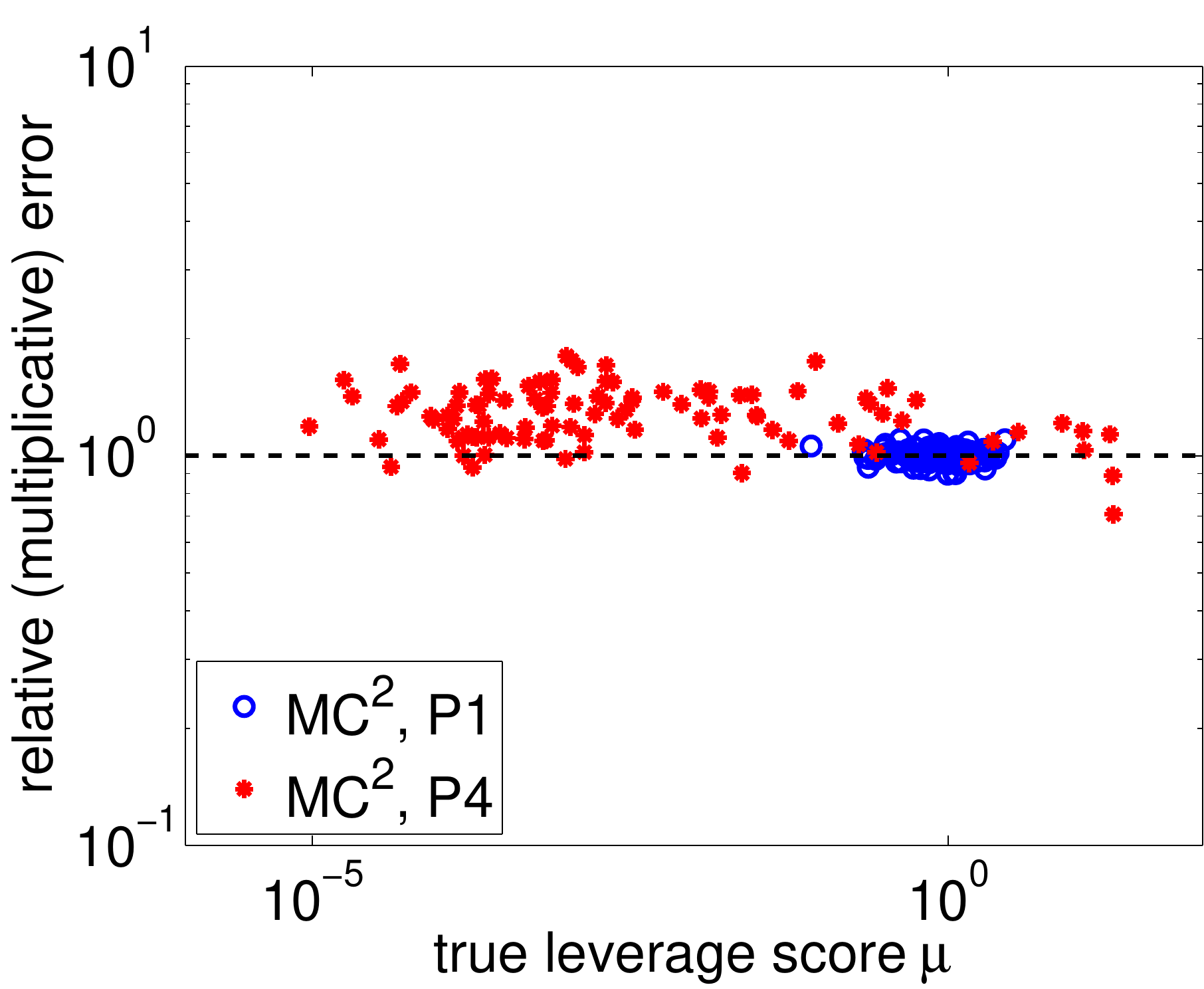} ~~~
(d) \includegraphics[width=2.2in]{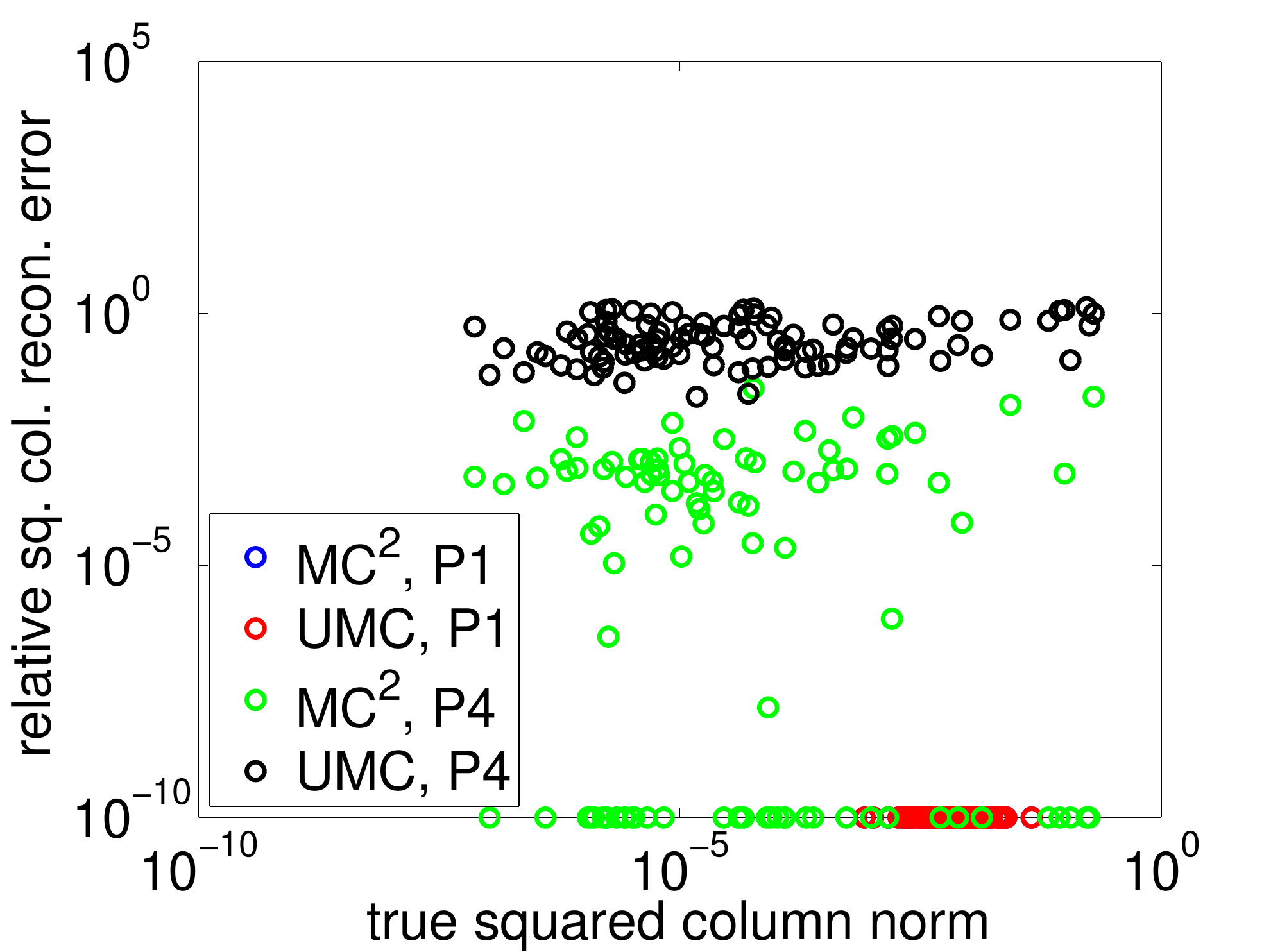}
\\
(e) \includegraphics[width=2.2in]{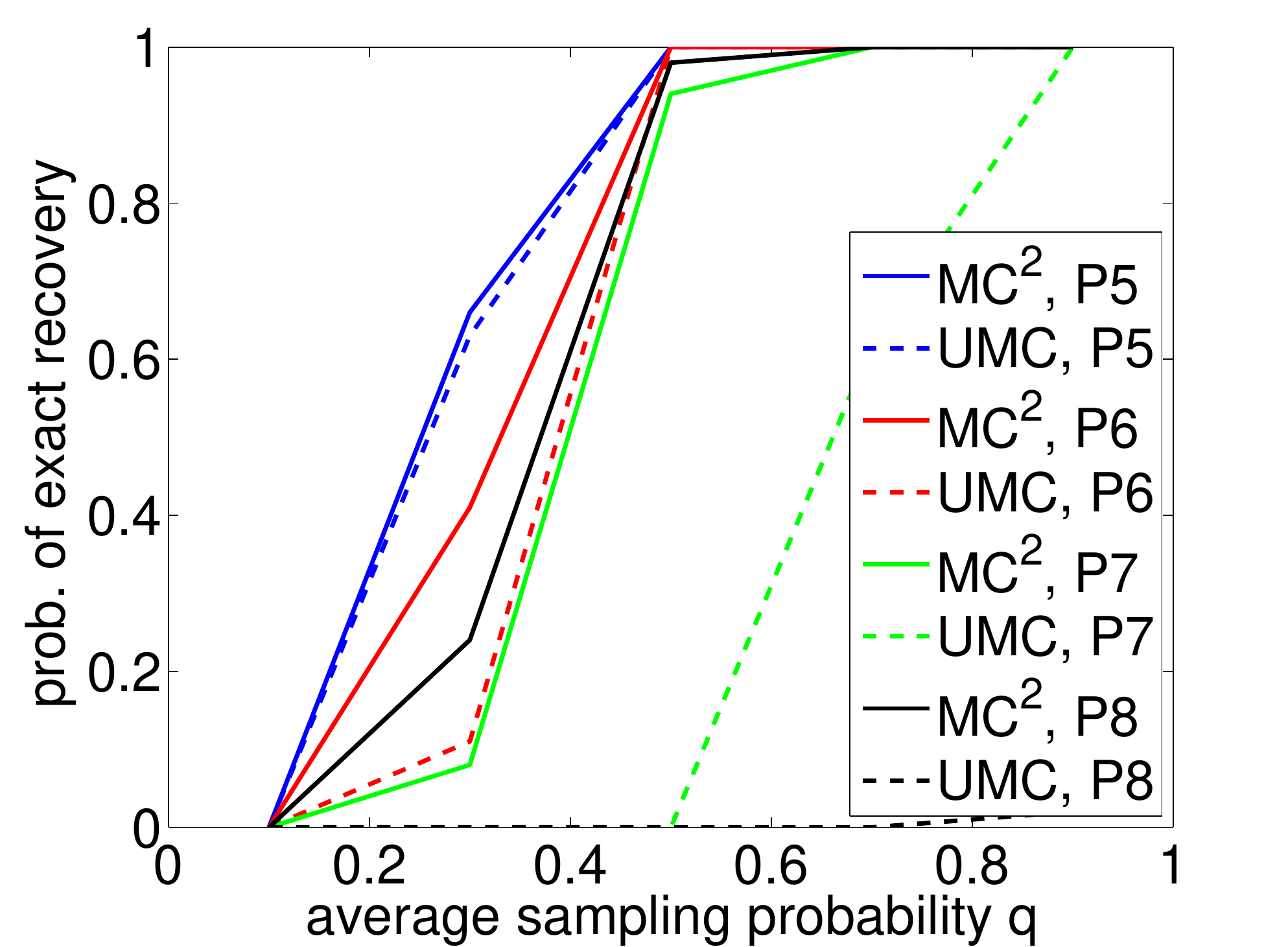} ~~~
(f) \includegraphics[width=2.2in]{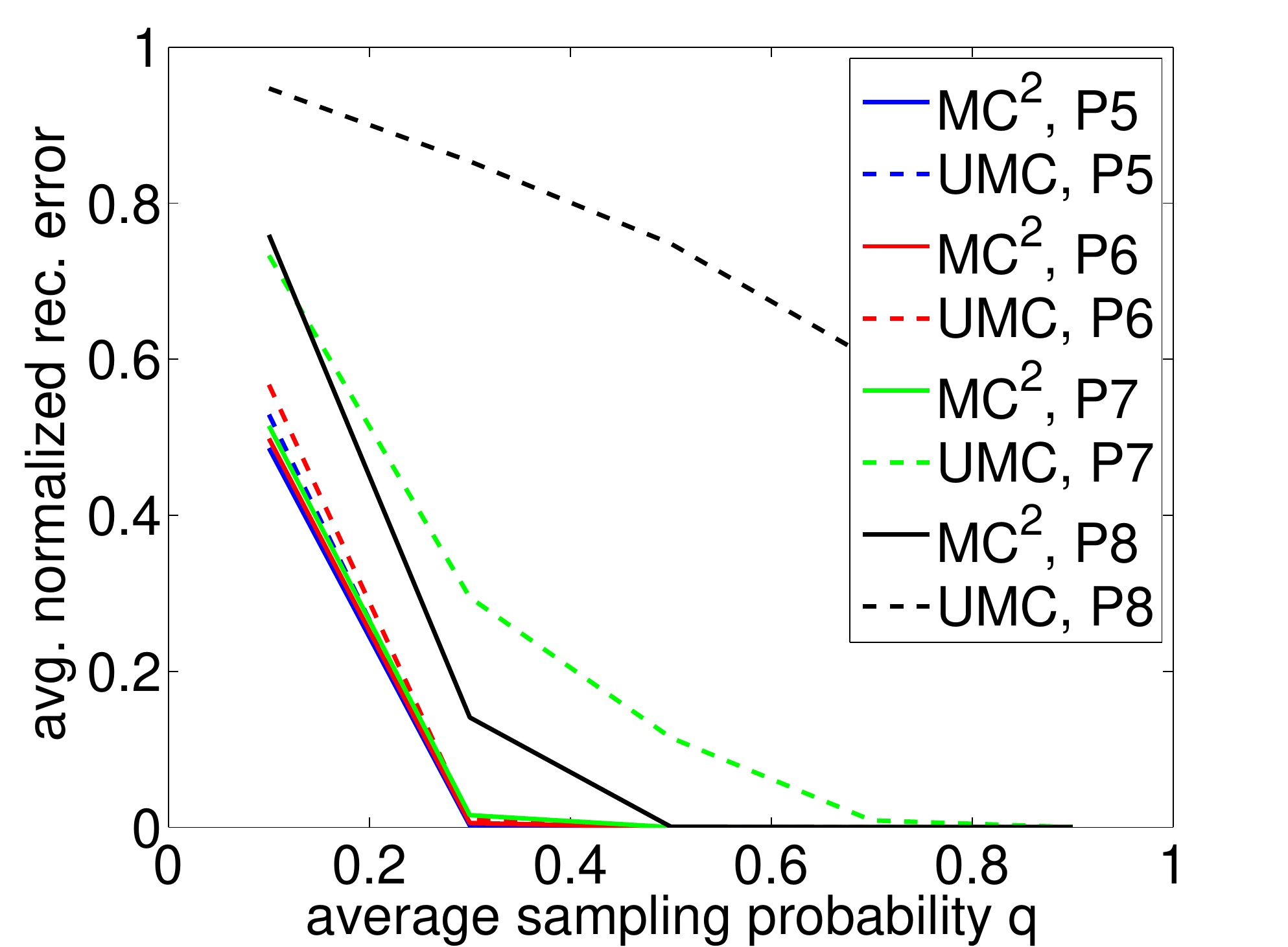}
\\
(g) \includegraphics[width=2.2in]{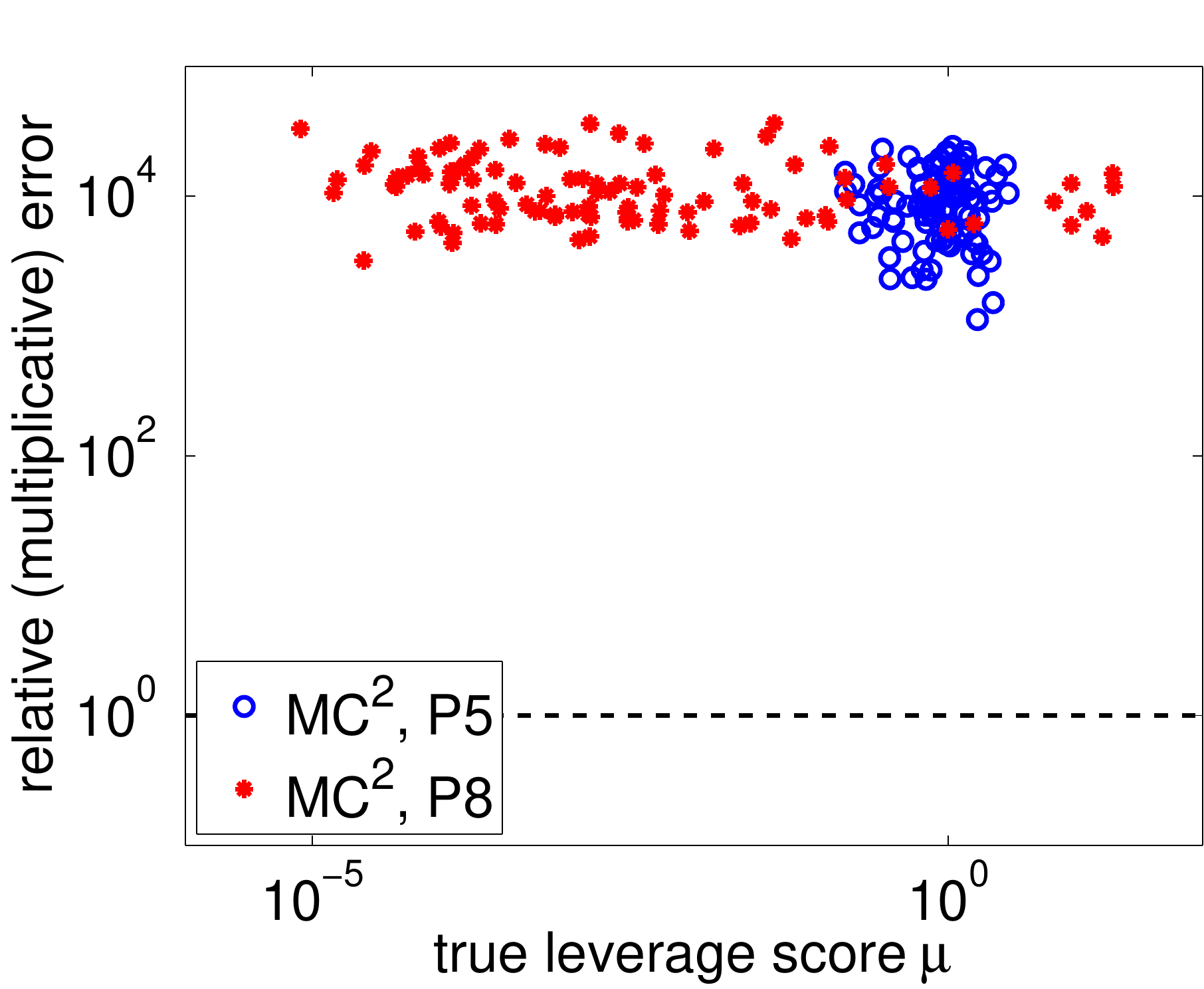} ~~~
(h) \includegraphics[width=2.2in]{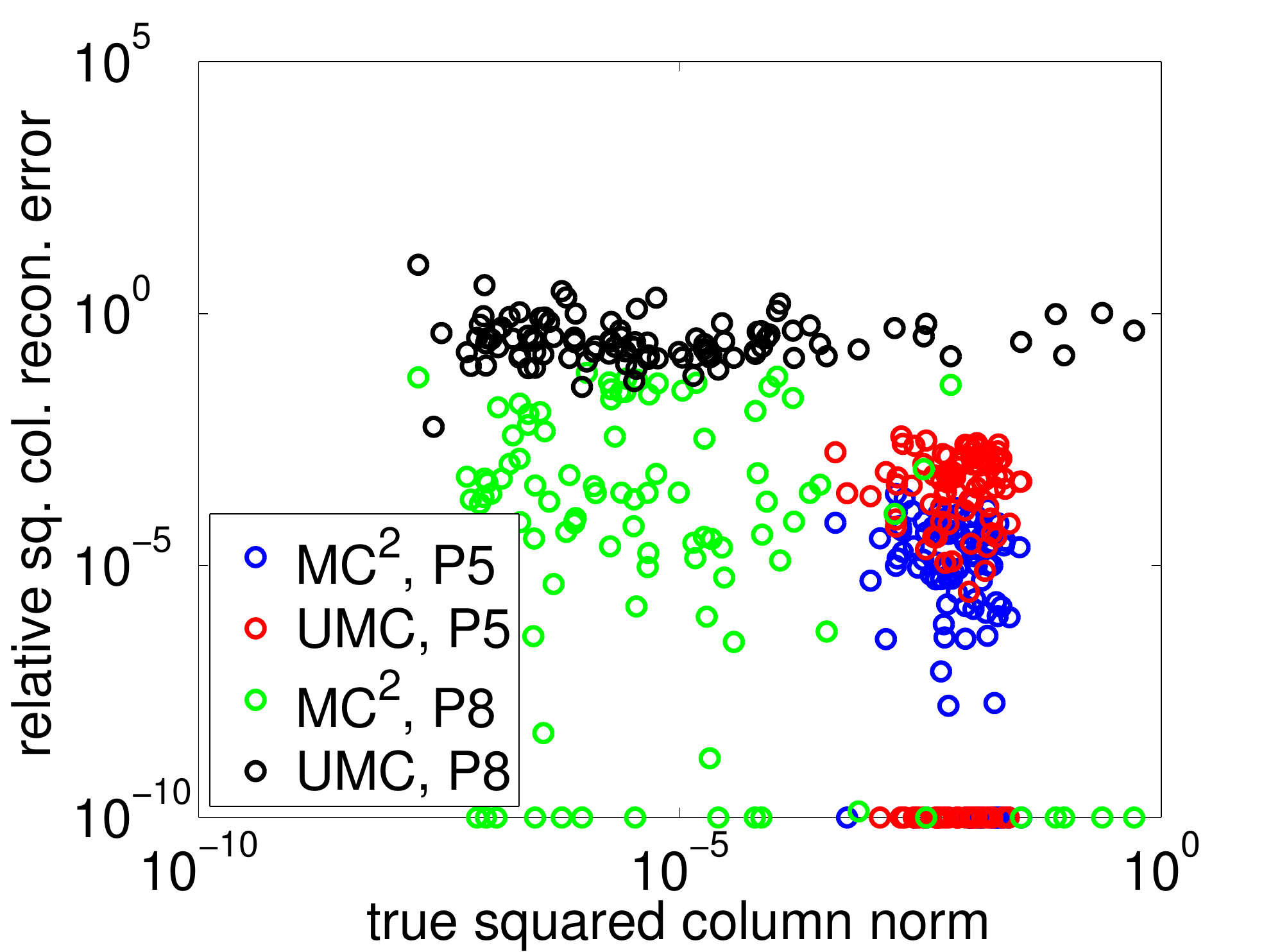}
\end{center}
\caption{\label{fig:power} Reconstruction results for power-law matrices. See Section~\ref{sec:powerdiscussion} for discussion.}
\end{figure}

Figures~\ref{fig:power}(e)--(h) show the analogous results for the poorly-conditioned power-law matrices P5--P8. Once again, we see that $\alg$ performs better than UMC. We also see once again that the performance of both $\alg$ and UMC generally worsens as the coherence $\eta$ increases (moving from P5 toward P8). However, while the degradation in performance is again quite dramatic for UMC, it is only minor with $\alg$. From Figure~\ref{fig:power}(g), we see that the average leverage score estimates are generally much larger than the true leverage scores (by an amount of approximately $10^4 = \kappa^2$). This is consistent with the dependence on the condition number suggested by the upper bound in Lemma~\ref{lemma:estLev}, although it may be possible to reduce the $\kappa^4$ term to $\kappa^2$.

\subsection{Results on block diagonal matrices}
\label{sec:blockdiscussion}

\revise{Figures~\ref{fig:block}(a)--(d)} show the reconstruction/estimation results for the well-conditioned block diagonal matrices B1 and B2. We see that UMC outperforms $\alg$ on matrix B1, which is incoherent, but that $\alg$ outperforms UMC on matrix B2, which is coherent. All leverage score estimates in $\alg$ are again quite accurate, with multiplicative errors clustered around 1. We note that, in matrix B1, all true leverage scores are equal to $1$, and all true squared column norms $\|M[:,j]\|_F^2 = 10^{-2}$. On matrix B2, the largest energy columns correspond to the smallest blocks in the block diagonal matrix. UMC performs poorly in reconstructing these columns, while $\alg$ performs well.

\begin{figure}[p]
\begin{center}
(a) \includegraphics[width=2.2in]{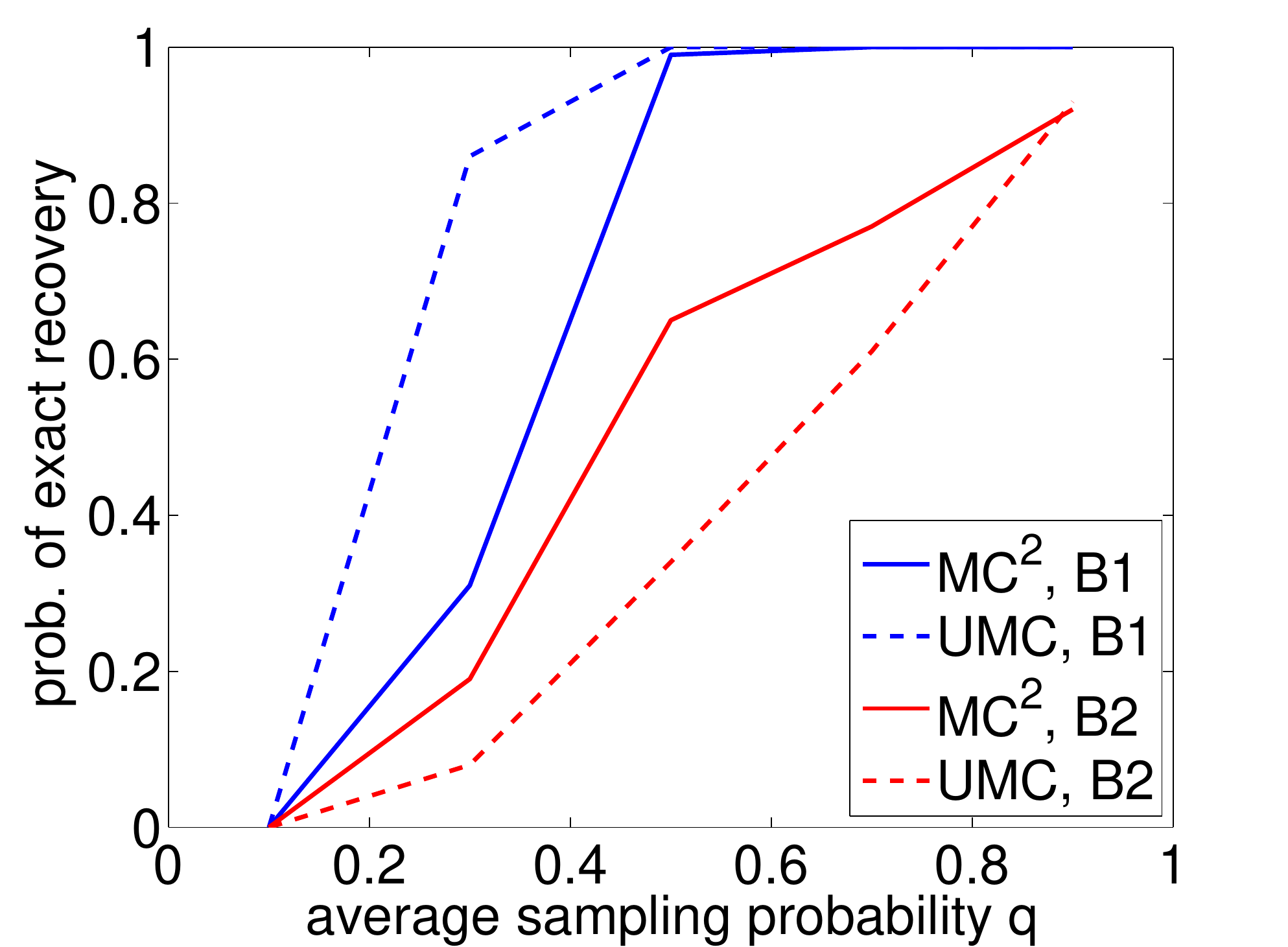} ~~~
(b) \includegraphics[width=2.2in]{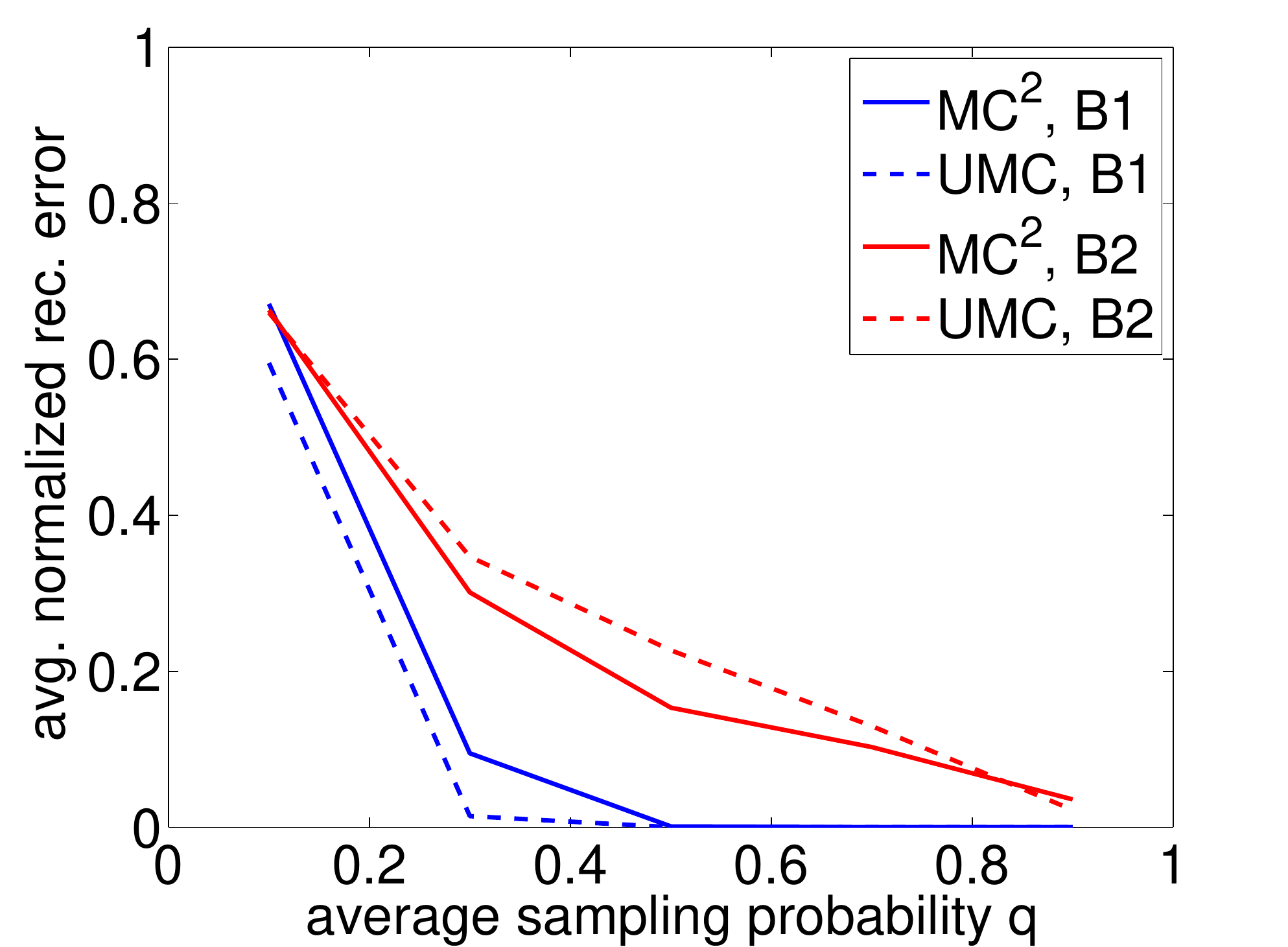}
\\
(c) \includegraphics[width=2.2in]{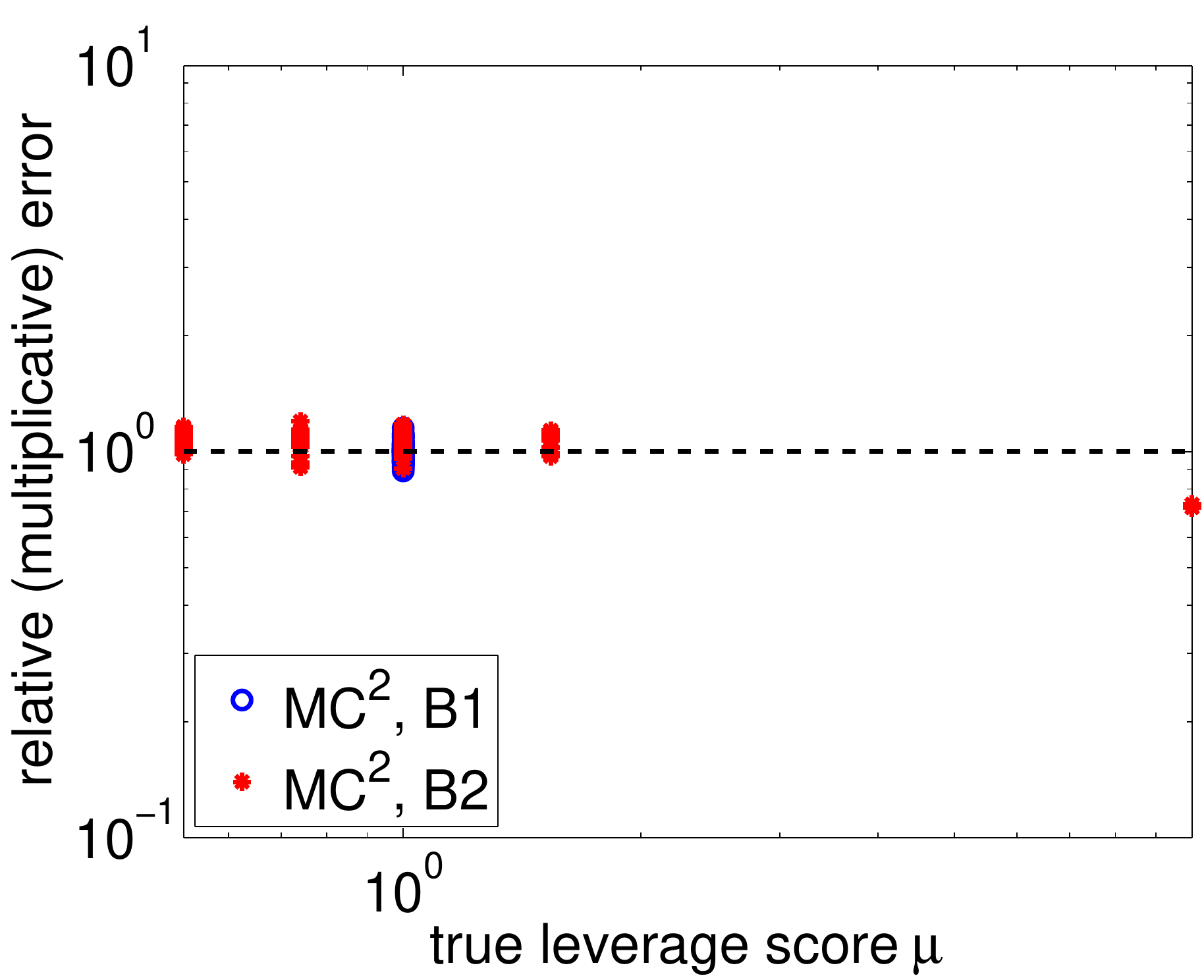} ~~~
(d) \includegraphics[width=2.2in]{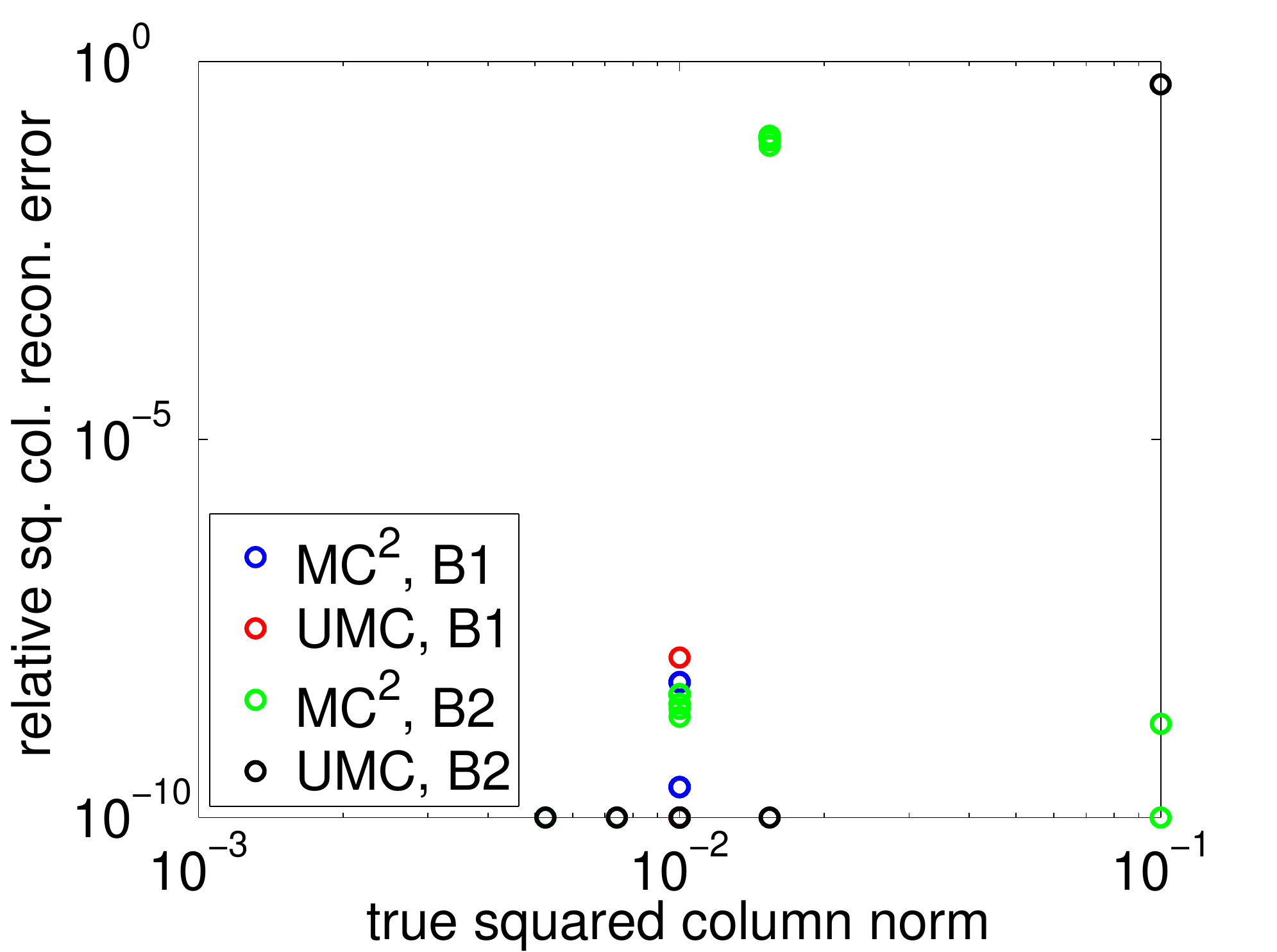}
\\
(e) \includegraphics[width=2.2in]{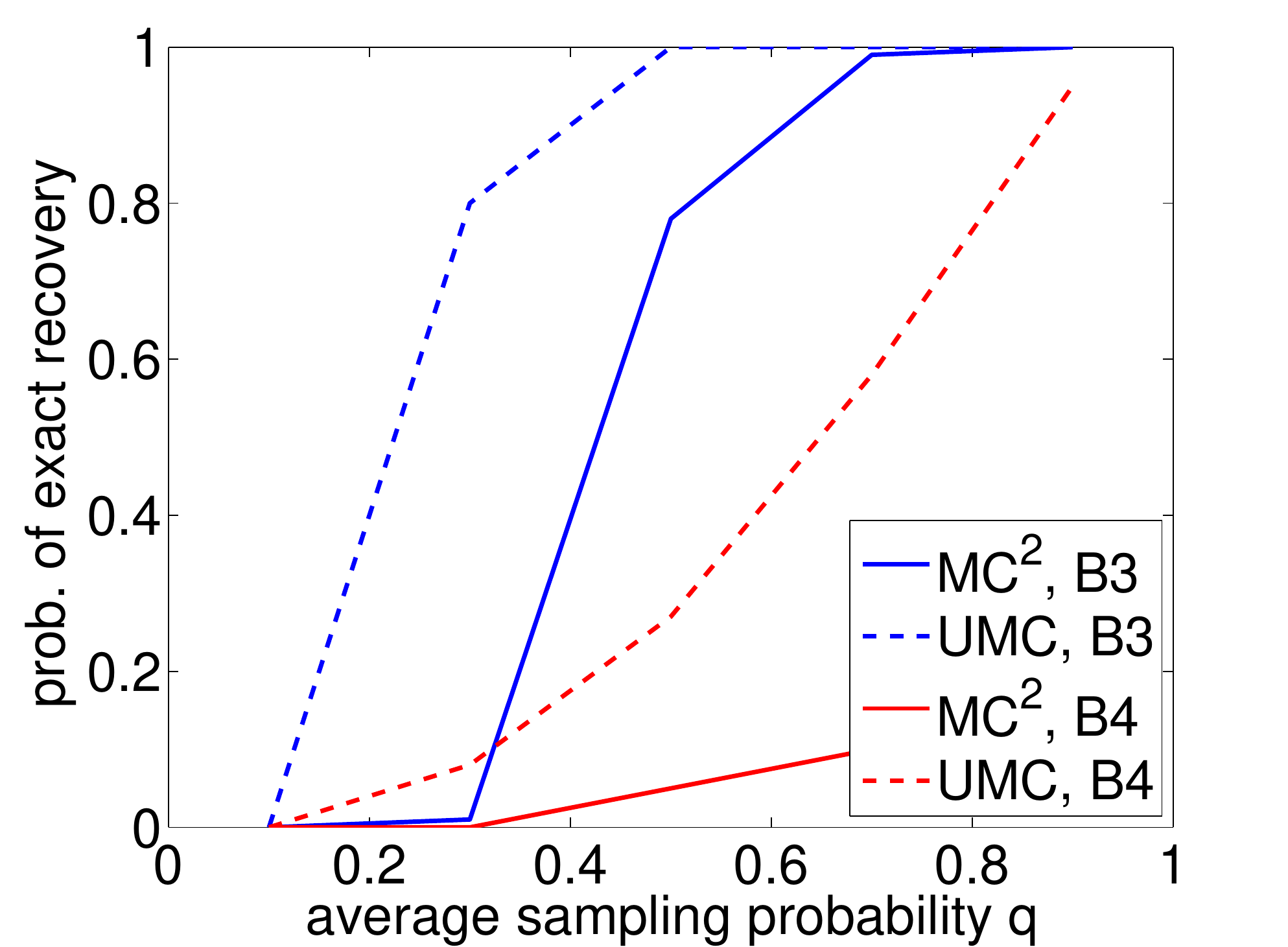} ~~~
(f) \includegraphics[width=2.2in]{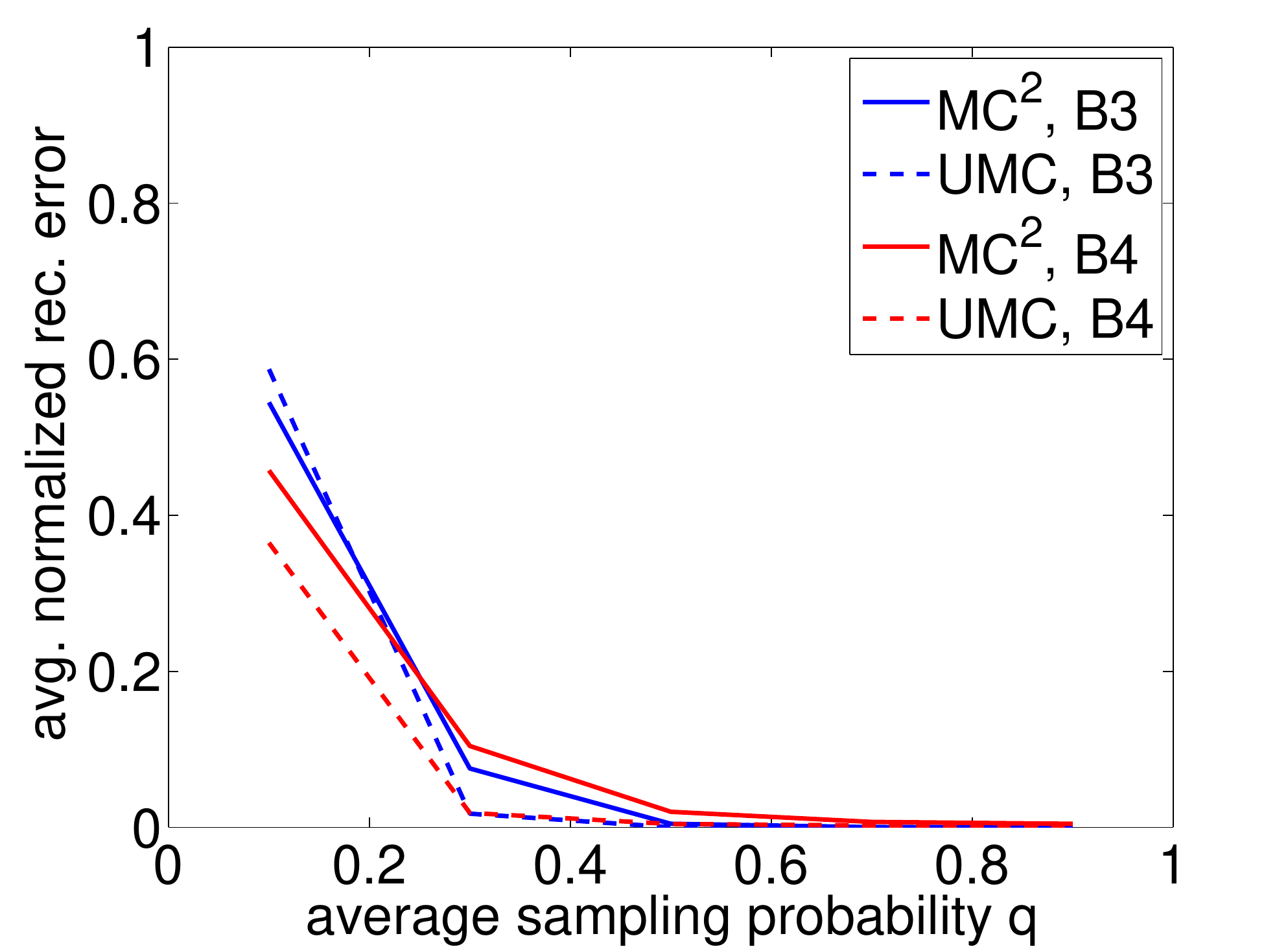}
\\
(g) \includegraphics[width=2.2in]{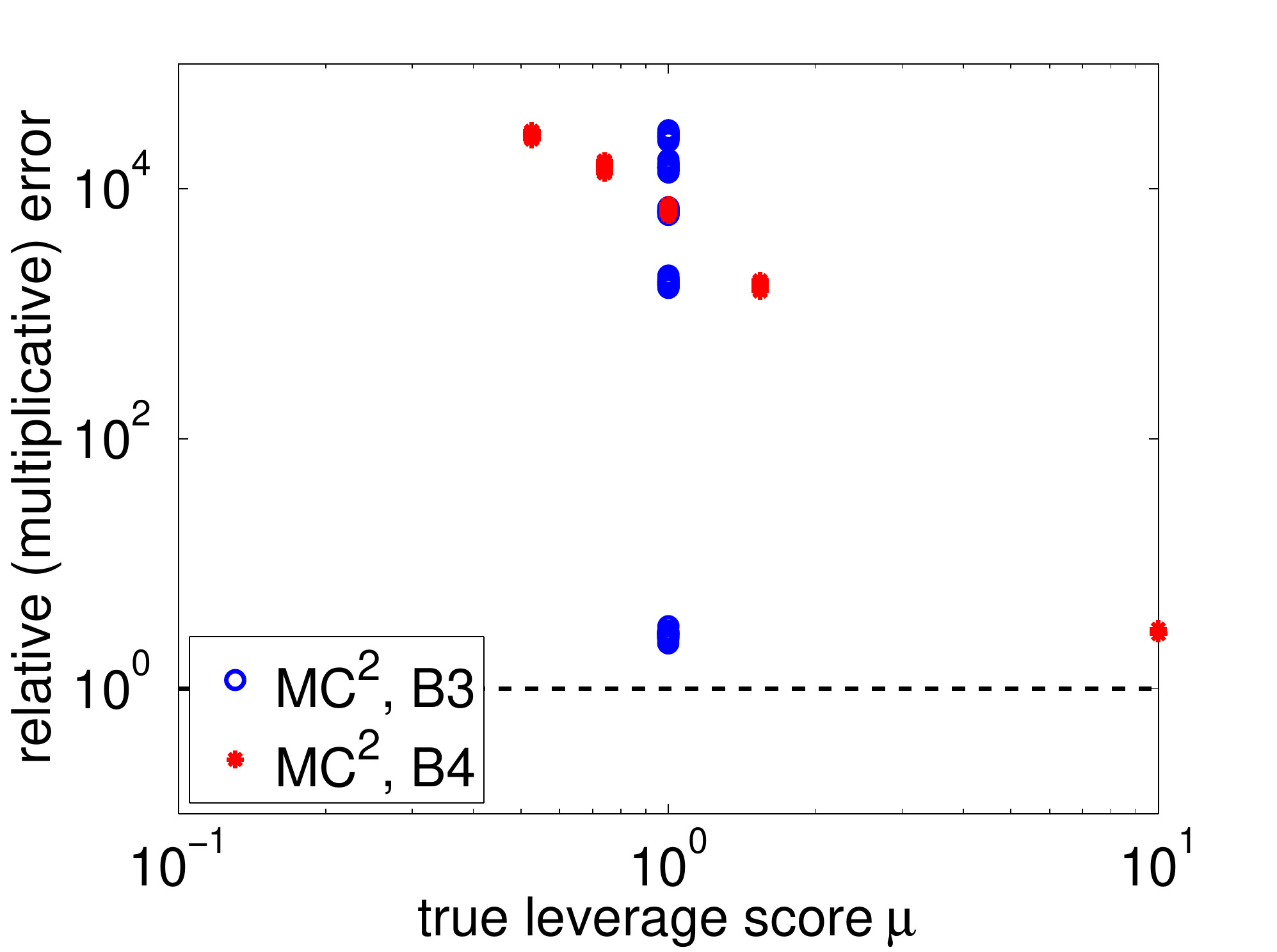} ~~~
(h) \includegraphics[width=2.2in]{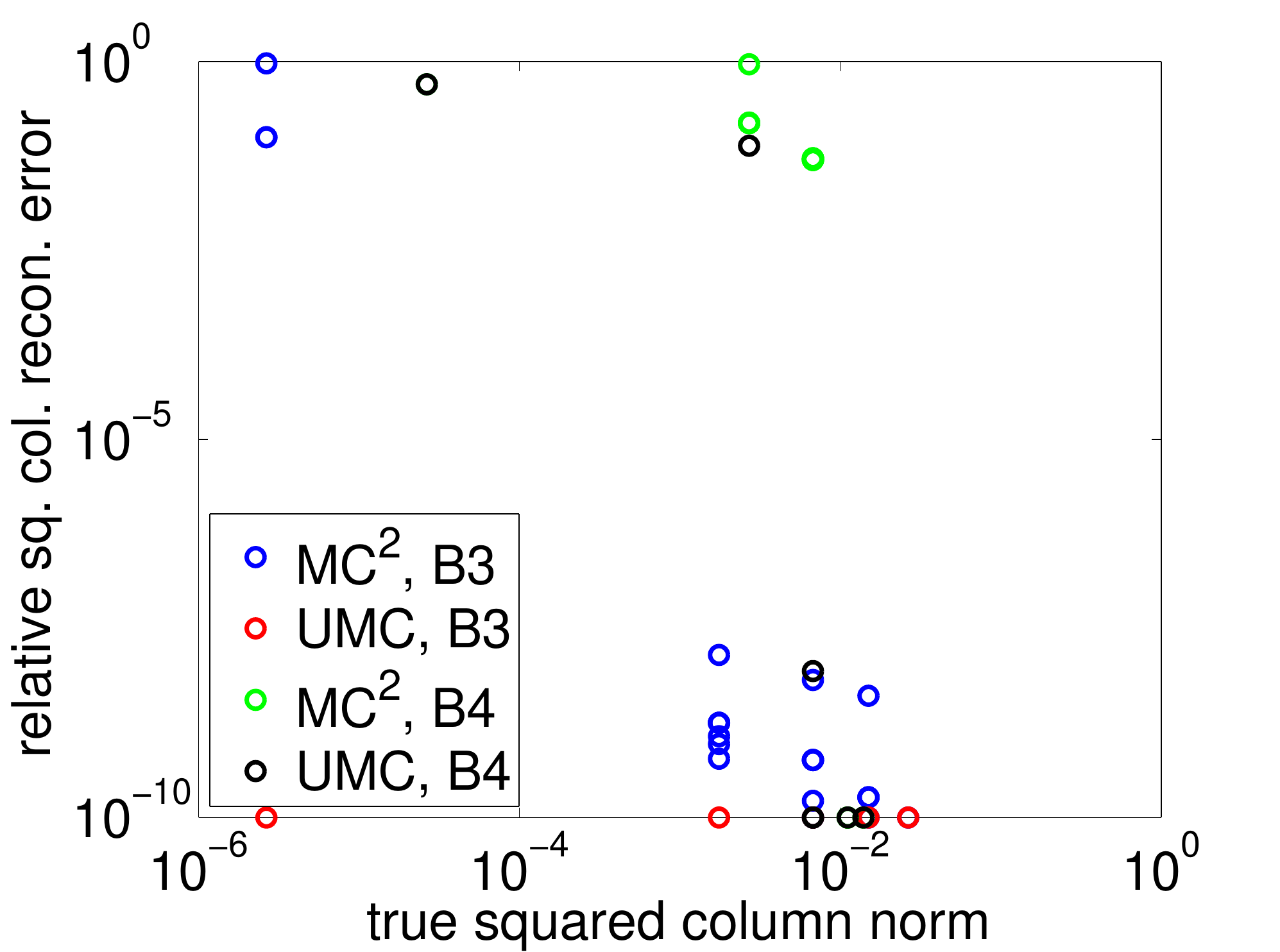}
\end{center}
\caption{\label{fig:block} Reconstruction results for block diagonal matrices. See Section~\ref{sec:blockdiscussion} for discussion.}
\end{figure}

\revise{Figures~\ref{fig:block}(e)--(h)} show the reconstruction/estimation results for the poorly-conditioned block diagonal matrices B3 and B4. These experiments illustrate the potential limitations of $\alg$ on poorly-conditioned matrices, where inaccurate estimates of leverage scores can lead to poor reconstruction performance. On both matrices, UMC generally outperforms $\alg$. Moveover, we see that as with the poorly-conditioned power-law matrices, the multiplicative leverage score errors can be large, on the order of $10^4 = \kappa^2$. On matrix B3, all true leverage scores are equal to $1$. Interestingly, on matrix B4, the larger leverage scores (which correspond to the smaller blocks in the block diagonal matrix) are estimated more accurately. We also see a general downward trend in the column reconstruction errors; with some exceptions, the larger norm columns (which correspond to the bottom right blocks in matrices B3 and B4) are recovered more accurately.

\subsection{Results with noise}
\label{sec:noisediscussion}

Finally, we repeat the reconstruction experiments with noisy samples. Recalling that all matrices $M$ are normalized such that $\|M\|_F = 1$, we add i.i.d.\ Gaussian noise to the samples with mean 0 and standard deviation $\sigma$. We modify the reconstruction steps in UMC and $\alg$ by replacing the equality constraint with an inequality constraint $\| P_\Omega(X - M) \|^2_F \le \delta$, where $\delta$ is set as an oracle according to the true noise energy on the set $\Omega$. Reconstruction errors are shown in \revise{Figures~\ref{fig:noise}(a)--(e)} for the power-law matrices P1, P4, and P8 and block diagonal matrices B2 and B4, respectively. We set the noise parameter $\sigma = 0.001$ (blue curves) or $\sigma = 0.01$ (red curves). In general, the performance of both $\alg$ (solid curves) and UMC (dashed curves) worsens as the noise level increases. As in the noise-free experiments, the performance of $\alg$ and UMC are quite close on matrix P1; $\alg$ generally outperforms UMC on matrices P4, P8, and B2; and UMC outperforms $\alg$ on matrix B4.

\begin{figure}[p]
\begin{center}
(a) \includegraphics[width=2.2in]{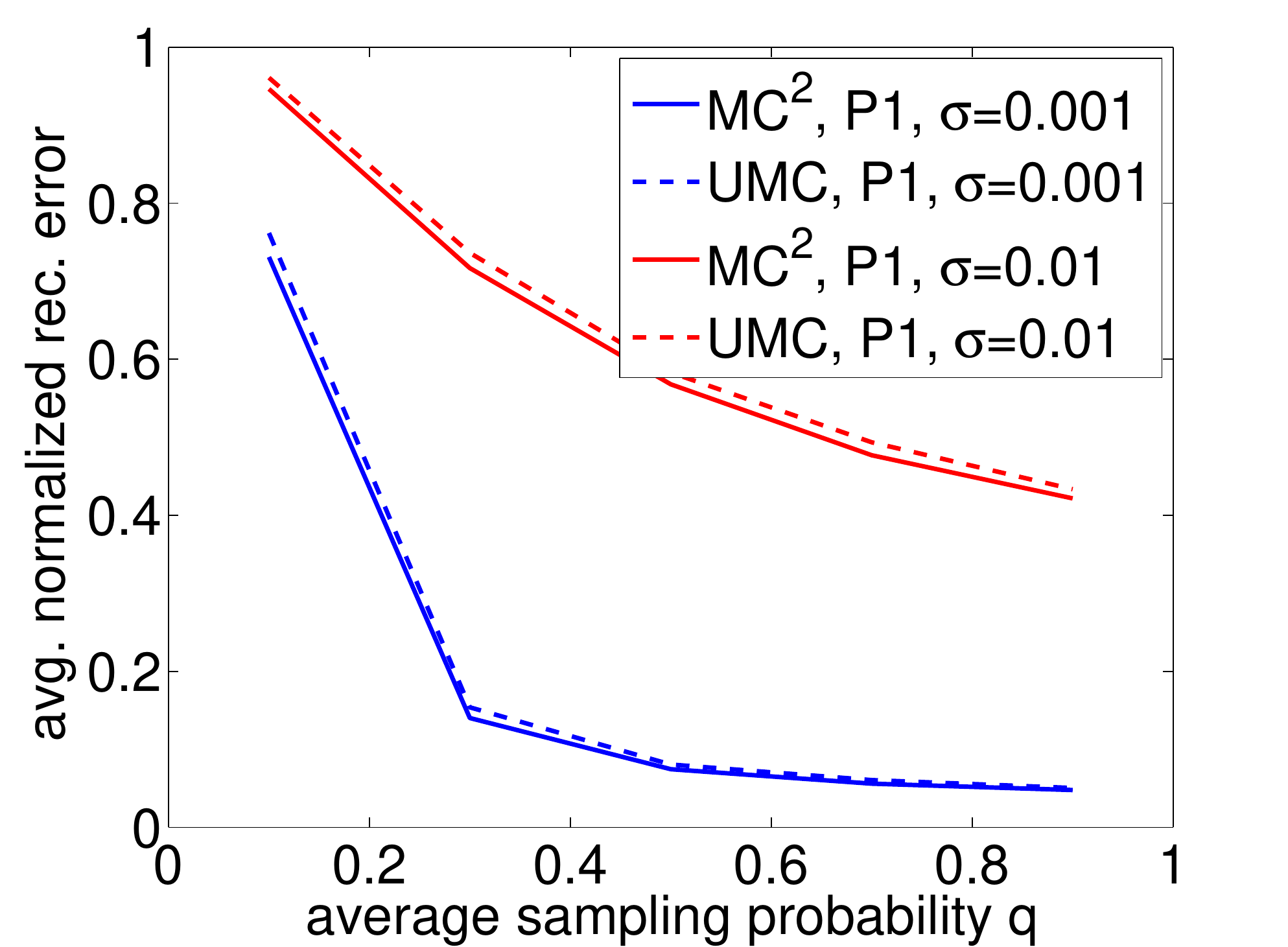} ~~~
(b) \includegraphics[width=2.2in]{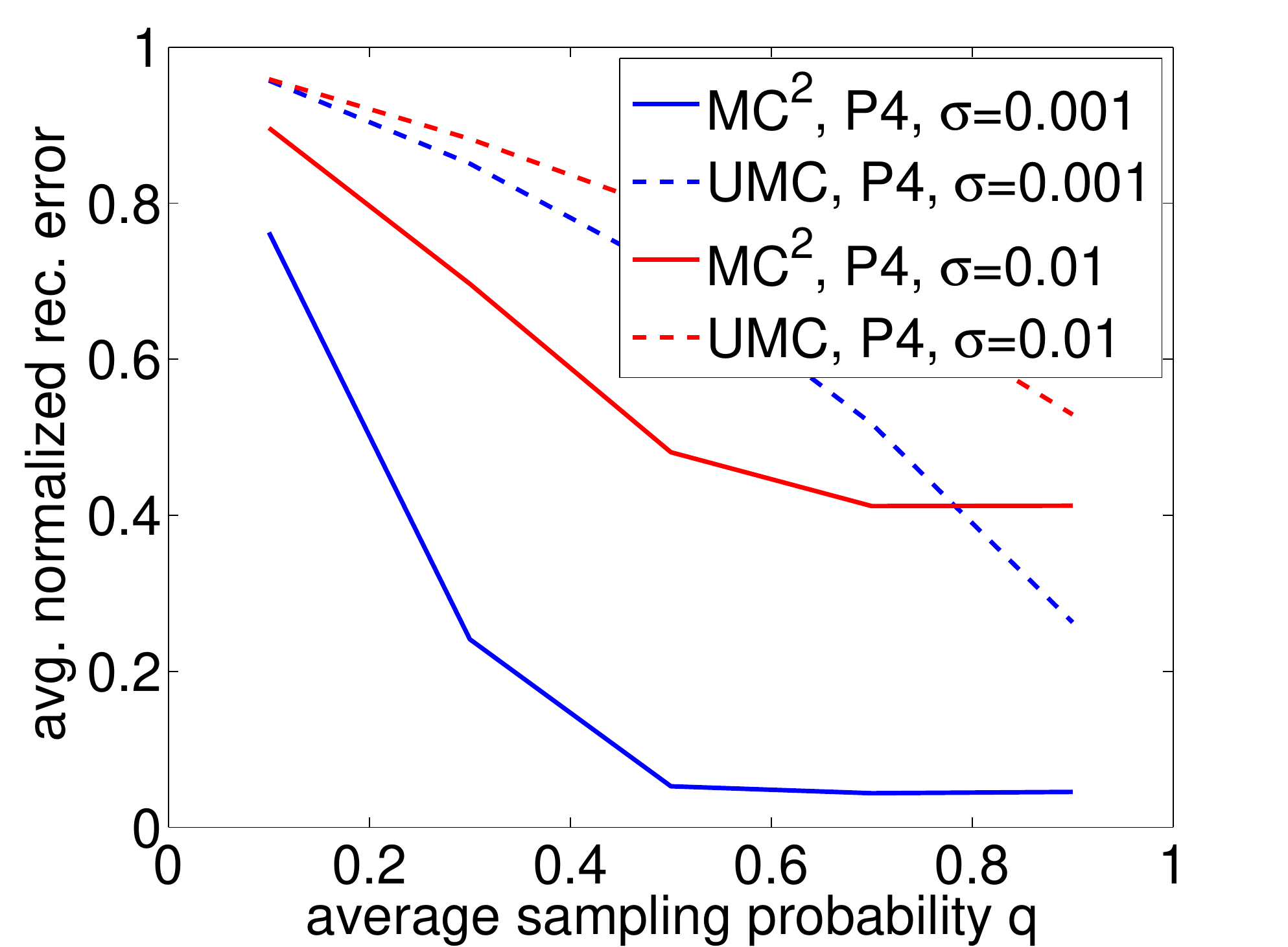}
\\
(c) \includegraphics[width=2.2in]{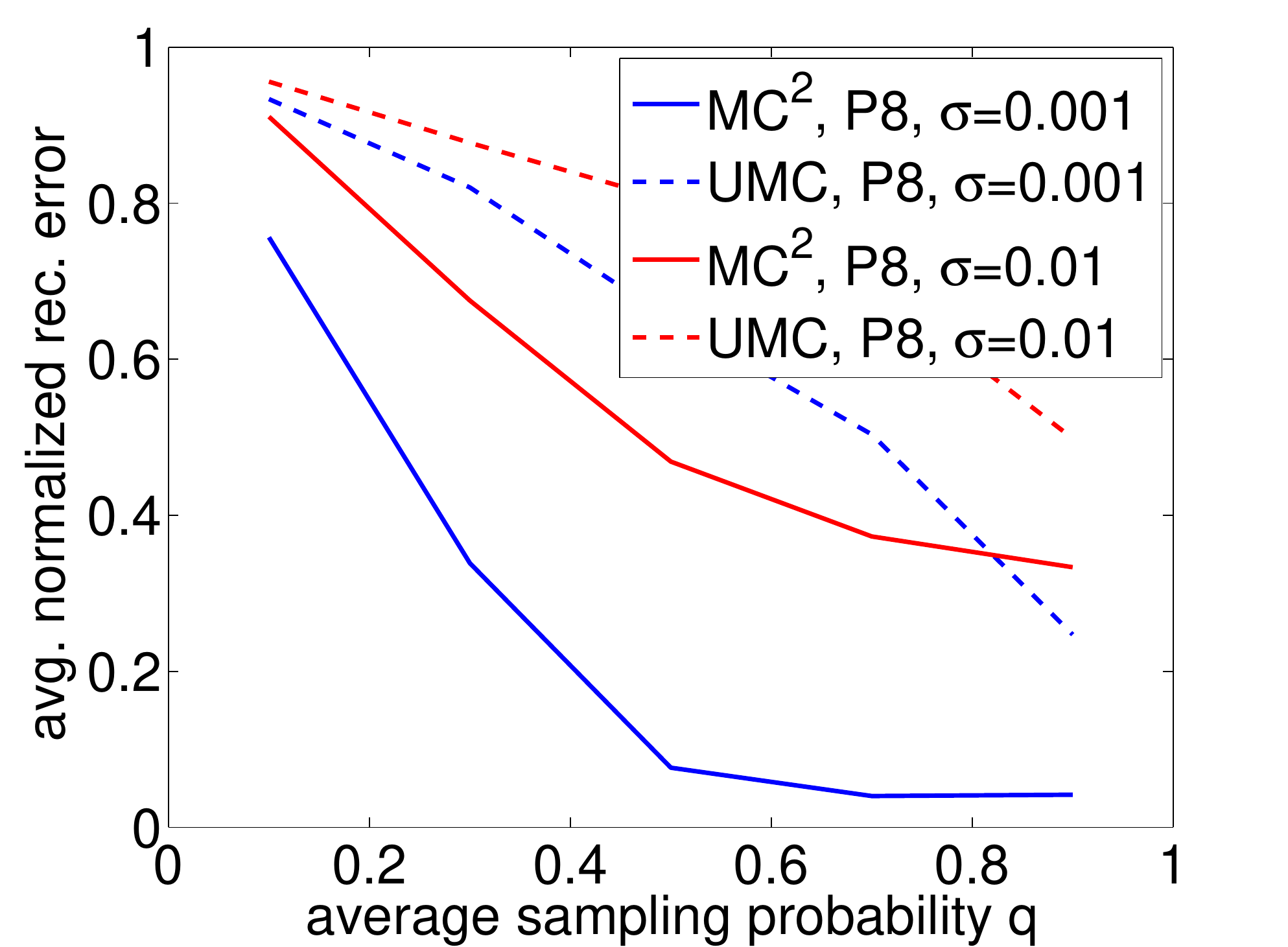} ~~~
(d) \includegraphics[width=2.2in]{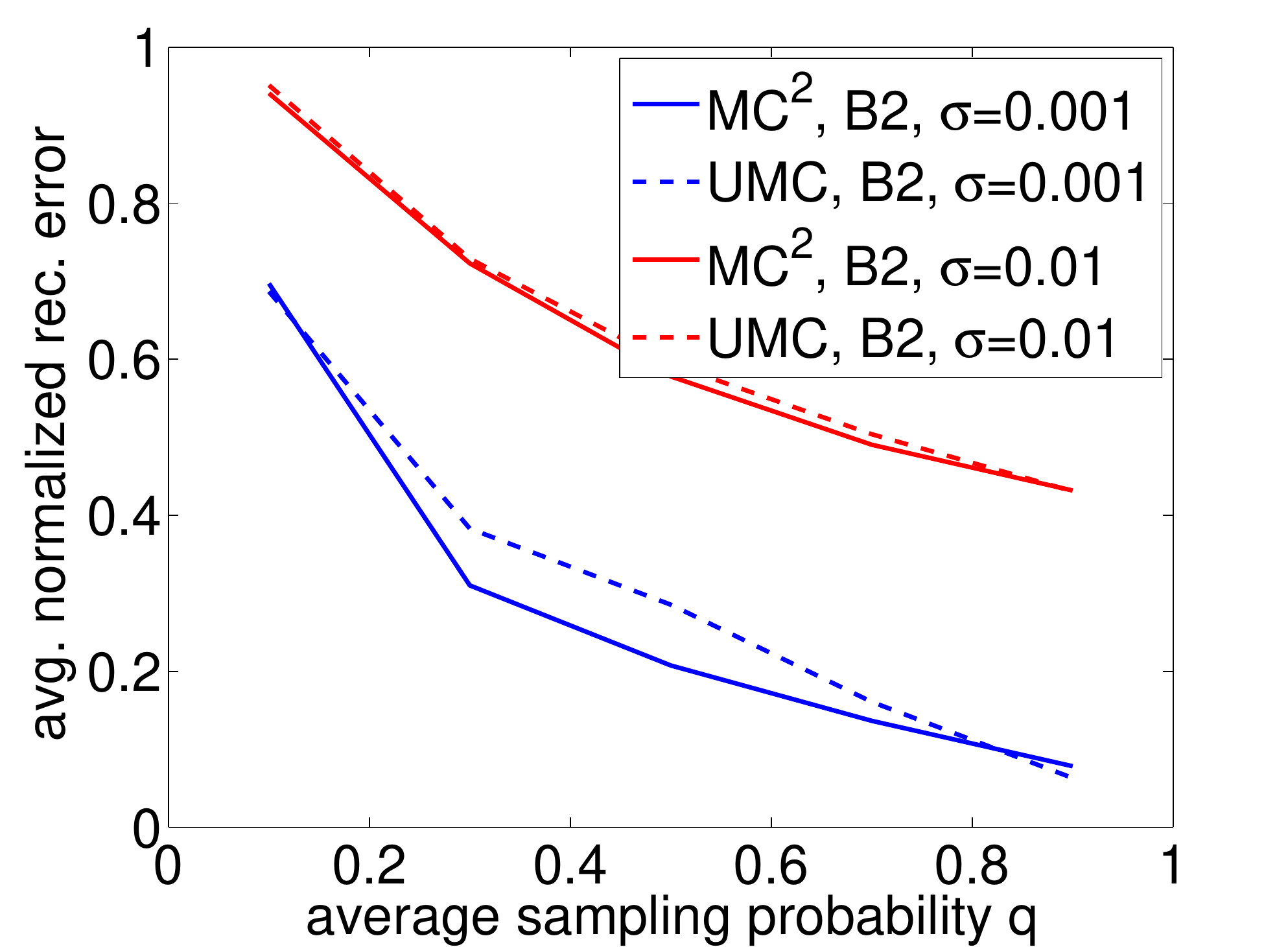}
\\
(e) \includegraphics[width=2.2in]{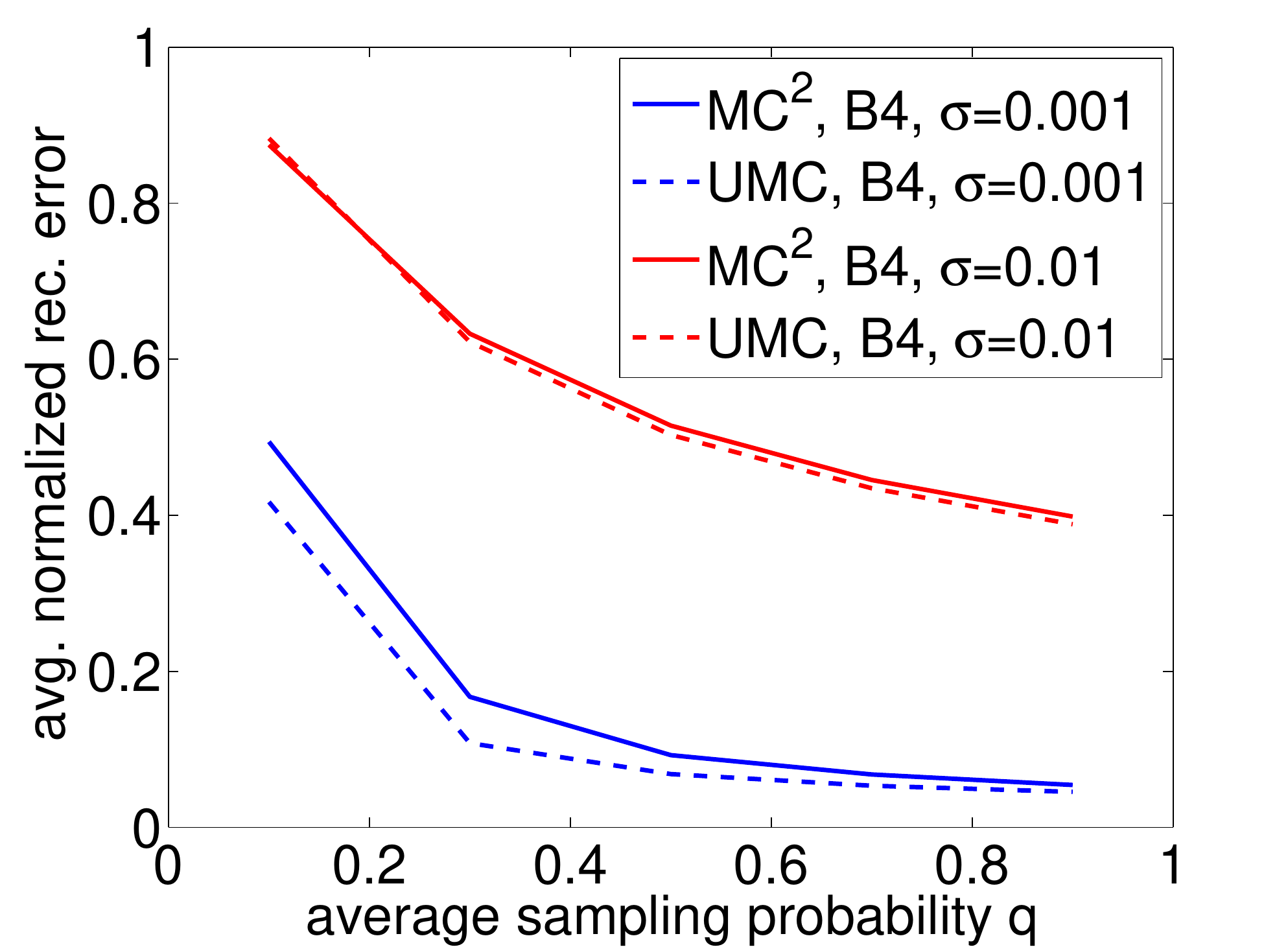}
\end{center}
\caption{\label{fig:noise} Reconstruction results with noisy samples. See Section~\ref{sec:noisediscussion} for discussion.}
\end{figure}

\section{Related Work} \label{sec:Related-Work}

Matrix completion is an active research topic with a myriad of  practical applications, and the large body of related literature includes  \cite{candes2009exact,candes2010power,recht2011simpler,candes2010matrix,keshavan2009matrix,jain2013low,chen2013completing,
negahban2012restricted,laurent2009matrix,davenport20141,eriksson2012high,chen2015incoherence}. An extension of the ideas in \cite{chen2013completing} to tensor completion recently appeared in \cite{bhojanapalli2015new}.

Leverage scores are of particular interest in numerical linear algebra. For instance, in a large linear regression problem, working with a randomly-selected small row-subset of the design matrix will significantly improve the processing time, without adversely affecting the performance. On this front, a few  relevant references are  \cite{cohen2015uniform,halko2011finding,meng2013low,mohri2011can,holodnak2015conditioning,drineas2012fast},
 where estimation of leverage scores is discussed. It is worth pointing out that, rather than row and column norms used in Theorem~\ref{thm:main}, one may alternatively utilize the column and row subspaces of $Y$ (the measurement matrix), after truncation,  to estimate the leverage scores. This, however, evidently leads  to an additive (rather than multiplicative)
error bounds (akin to \cite{holodnak2015conditioning}) which are  in fact not suited for the analysis of $\alg$ here.

\section*{Acknowledgments}

The authors thank Dehui Yang at Colorado School of Mines for helpful discussions in early phases of this project. Part of this research was conducted when AE was a visitor at  the Institute for Computational and Experimental Research in Mathematics (ICERM). AE is grateful for their hospitality and kindness.  MBW was partially supported by NSF CAREER Grant CCF-$1149225$ and NSF Grant CCF-$1409258$. RW was partially funded by NSF CAREER Grant  CCF-$1255631$. Finally, the authors thank the anonymous referees for their detailed comments and helpful suggestions, including those for the experimental framework.

\bibliographystyle{plain}
\bibliography{References}

\end{document}